\documentclass[3p,times]{elsarticle}

\usepackage{amsfonts}
\usepackage{graphicx}
\usepackage{epstopdf}
\usepackage{lineno}
\usepackage[utf8]{inputenc} % allow utf-8 input
\usepackage[T1]{fontenc}    % use 8-bit T1 fonts
\usepackage{hyperref}       % hyperlinks
\usepackage{url}            % simple URL typesetting
\usepackage{multirow}
\usepackage{siunitx}
\usepackage{nicefrac}       % compact symbols for 1/2, etc.
\usepackage{microtype}      % microtypography
\usepackage{mathtools}      % loads amsmath as well
\usepackage{commath}      % loads amsmath as well
\usepackage{amssymb}	    % extra math symbols
\usepackage[english]{babel}
\usepackage{float}
\usepackage{bm}
\usepackage{indentfirst}
\usepackage{tikz}
\usepackage{pgfplots}
\usepackage{algorithmic}
\usepackage{color,soul}
\usepackage{algorithm2e}
\usepackage{enumitem}
\usepackage{rotating}
\usepackage{caption}
\usepackage{subcaption}
\usepackage[subtle]{savetrees}
\usepackage{booktabs}
\ifpdf
  \DeclareGraphicsExtensions{.eps,.pdf,.png,.jpg}
\else
  \DeclareGraphicsExtensions{.eps}
\fi
\usepackage{cleveref}       % clever cross-referencing

\crefname{equation}{}{}
\crefname{appendix}{}{}

\newcommand{\FLASH}{{\sc flash}}

\newcommand{\model}[1]{\texttt{#1}}
\newcommand*{\I}{\imath}

\newcommand{\comment}[1]{} % create block comments

\definecolor{mygray}{RGB}{145,145,145}
\definecolor{myred}{RGB}{255,60,60}
\definecolor{myblue}{RGB}{60,60,255}

\pgfplotsset{width=7cm,compat=1.10}
\usetikzlibrary{shapes,arrows,fit,calc,positioning}
\usetikzlibrary{decorations.markings}
\modulolinenumbers[5]

\journal{Computer Physics Communications}

\begin{document}

\begin{frontmatter}

\title{A hybrid adaptive multiresolution approach for the efficient simulation
    of reactive flows}

\author[fsu]{Brandon Gusto\corref{cor1}}
\ead{blg13@my.fsu.edu}

\author[fsu]{Tomasz Plewa}
\ead{tplewa@fsu.edu}

\cortext[cor1]{Corresponding author}

\address[fsu]{Department of Scientific Computing, Florida State University, 600 W College Ave,
    Tallahassee, FL 32306, U.S.A}

\begin{abstract}
    Computational studies that use block-structured adaptive mesh refinement
    (AMR) approaches suffer from unnecessarily high mesh resolution in regions
    adjacent to important solution features. This deficiency limits the
    performance of AMR codes. In this work a novel hybrid adaptive
    multiresolution (HAMR) approach to AMR-based calculations is introduced to
    address this issue. The multiresolution (MR) smoothness indicators are used
    to identify regions of smoothness on the mesh where the computational cost
    of individual physics solvers may be decreased by replacing direct
    calculations with interpolation. We suggest an approach to balance the
    errors due to the adaptive discretization and the interpolation of physics
    quantities such that the overall accuracy of the HAMR solution is consistent
    with that of the MR-driven AMR solution. The performance of the HAMR scheme
    is evaluated for a range of test problems, from pure hydrodynamics to
    turbulent combustion.
\end{abstract}


\end{frontmatter}
\section{Introduction}
\label{sec:introduction}
%
%
%
    % overview paragraphs
    Combustion is a physical process that plays a critical role in many
    industrial, defense, and basic science applications. Oftentimes combustion
    is accompanied by the chaotic mixing process known as turbulence. These
    phenomena are prevalent in problems ranging from internal combustion
    engines to stellar evolution. Significant progress has been made in the
    numerical modeling of such processes (c.f.\ \cite{chen2011}). However, as the
    physics models used to describe turbulent combustion become increasingly
    complex, the need for more efficient computational tools increases
    commensurately.

    One of the major challenges associated with simulations of turbulent
    combustion is the presence of disparate length and time scales.  In particular,
    the thermonuclear burning timescale in the problem of detonation ignition
    and turbulent stellar plasma \cite{fenn2017} is on the order of
    $\tau_{\mathrm{burn}} \approx 1 \times 10^{-9}$ s, whereas the eddy turnover
    time is $\tau_{\mathrm{turb}} \approx 1 \times  10^{-2}$ s.  At the same
    time, the turbulent kinetic energy transfer couples the turbulence driving
    scales, $l_{\mathrm{td}} \approx 1 \times 10^{6}$ cm, with the viscous
    dissipation scale, which is several orders of magnitude smaller.  This
    disparity in scales frequently makes direct numerical simulations
    computationally infeasible, forcing researchers to reduce the range of
    scales by introducing approximate descriptions of physics operating on
    unresolvable scales \cite{oran+00}.  The above situation might be
    further complicated by the contribution of additional effects such as
    compressibility (due to coupling to burning) and anisotropy (due to
    gravity). In the case of compressibility, provided that the resulting energy
    release is sufficiently high, acoustic perturbations might be strengthened
    into shocks and ultimately become detonation waves \cite{zeldovich+70}. The
    reaction zones supporting detonations not only see extreme changes in
    internal energy, but are also small in comparison to the size of the
    computational domain.  Furthermore, the physical description of the
    participating medium requires in this case the use of a complex stellar
    equation of state (EOS). This type of EoS is very costly to evaluate in
    comparison to hydrodynamics \cite{timmes2000a}.

    % paragraph introducing adaptive mesh refinement as a concept
    The above problem description indicates that accurately resolving turbulence
    and reactive scales may not be possible when using a uniform spatial
    resolution.  One class of numerical discretization algorithms designed to
    address the challenge of disparate scales are adaptive mesh
    refinement (AMR) methods. These methods employ a hierarchical approach to
    progressively resolve solution structure, with the refinement process being driven
    by relevant mesh refinement criteria.  These criteria are of key importance
    in achieving the required solution accuracy at minimal computational cost.
    To adapt the mesh, Berger \& Oliger \cite{berger1984} used Richardson
    extrapolation to estimate the local truncation error (LTE) of the solution.
    Because the solution error is expected to depend on the amount of variation
    in the solution, less sophisticated methods that rely on the first-order
    solution variation \cite{quirk1991}, or possibly higher-order derivatives
    \cite{lohner1987} have been used for that purpose. In certain situations it
    might be useful to drive mesh refinement using application-specific
    refinement criteria, such as abundance or rate of mixing of reactive
    species. (For a partial review of the LTE estimators and other refinement criteria,
    as well as an analysis of AMR solution error for certain classes of partial
    differential equations (PDEs), see \cite{li2010}.)

    % paragraph reviews the work of harten and multiresolution methods
    Alternative approaches to dynamically adaptive schemes based on
    multiresolution analysis (MRA) have gained popularity in the computational
    fluid dynamics community over the last two decades \cite{schneider2010}.
    MRA provides a mathematically rigorous setting for the representation of
    discrete data in terms of nested approximation spaces, and serves as the
    basis of multiresolution (MR) adaptive techniques for solving PDEs
    \cite{harten1996}. The seminal papers by Harten \cite{harten1994,harten1995}
    introduced a MR scheme for solving hyperbolic conservation laws (HCLs) that
    reduces computational expense by adaptively computing the numerical flux. In
    Harten's scheme, the MR indicators are used to identify solution regions of
    sufficient smoothness where costly, high resolution numerical fluxes do
    not need to be computed directly to achieve the desired accuracy. Instead
    those fluxes are interpolated using high resolution flux values already
    obtained nearby.  In contrast to AMR methods, Harten's scheme evolves the
    solution uniformly at the finest level of mesh resolution.

    Initially, Harten's scheme was applied solely to the Euler equations in one
    spatial dimension, but was then extended by Bihari et~al.\ to include
    viscous effects \cite{bihari1996}, to two-dimensional situations
    \cite{bihari1997}, and also to PDEs with reactive source terms
    \cite{bihari1999}.  The same principle was also used by Chiavassa \& Donat
    \cite{chiavassa2001} to accelerate the computation of solutions to the
    multidimensional Euler equations. We refer to this class of MR schemes as
    \emph{solver adaptive}.

    % review the multiresolution-adaptive papers
    Although Harten's intent was to offer an algorithmically simpler alternative
    to grid adaptation \cite{harten1995}, the adaptive mesh setting was
    eventually used with the MR framework by Gottschlich-M\"uller \& M\"uller
    \cite{gottschlich-muller1999} to construct a fully adaptive MR scheme (see,
    also, \cite{kaibara2001}). Cohen et~al.\ \cite{cohen2003} also introduced a
    fully adaptive MR scheme for multidimensional HCLs, and provided rigorous
    error analysis. Roussel and collaborators used MR indicators to guide mesh
    adaptation for solving parabolic PDEs in \cite{roussel2003} and the reactive
    Euler equations in \cite{roussel2005,roussel2015}.  Domingues et~al.\
    \cite{domingues2008,domingues2009} combined this approach with locally
    adaptive time-stepping schemes. Generally speaking, these works employ AMR
    with the refinement criteria determined by the MR smoothness indicators.

    % quadtree/octree libraries:
    % - paramesh
    % - p4est
    % patch-based libraries
    % - amroc
    % - amrex
    % - chombo
    % - samrai
    In regard to the AMR mesh structure, it can be described as a collection of
    individual cells (cell-based format; \cite{khokhlov1998}), blocks containing
    a fixed number of cells (block-based, or block-structured format;
    \cite{powell1999}), or blocks containing varying number of cells
    (patch-based format; \cite{berger1984}). The efficiency of those various
    discretizations can be characterized using the mesh filling factor, which is
    the ratio of the number of cells required to resolve the solution in the AMR
    model to the number of cells in the corresponding uniformly resolved model.
    This mesh filling factor is the smallest possible for cell-based refinement,
    but usually higher for the latter two formats. However block- and
    patch-based formats are generally better suited for large distributed memory
    machines due to lower network communication costs (see, for example,
    \cite{gunney2016,vanstraalen2009,peplinski+16}).  The choice of which AMR
    discretization format to use depends on the computing hardware and
    application type.

    % introduce the work of peers
    Many libraries implementing the patch- or block-based AMR approaches are
    available. For example in our work we use the block-based PARAMESH library
    as implemented in the \FLASH\ code \cite{fryxell2000}. As mentioned earlier
    this approach allows for relatively simple mesh management with little
    associated metadata, and for simple physics applications scales well for
    problems requiring thousands of processors \cite{fisher+08}. The patch-based
    format is offered in numerous AMR libraries, including BoxLib/AMReX
    \cite{boxlib2011,amrex2019}, SAMRAI \cite{wissink2001}, AMROC
    \cite{deiterding2003}, and Chombo \cite{chombo2015}, among others.  Compared
    to the block-based format the patch-based refinement calculations may
    require more extensive metadata and more complex communication patterns,
    which may negatively impact load balancing in large-scale simulations. For
    these reasons, optimization of patch-based AMR algorithms remains an active
    area of research \cite{gunney2016,schornbaum2018}.

    Recently, MR approaches have been combined with block-structured AMR. For
    example, Han et~al.  \cite{han2014} employ a block-structured
    multiresolution adaptive approach for simulating multi-phase flows, with the
    blocks being organized in a pyramid-like data structure that allows overlap
    between neighbors. Sroka et~al.\ \cite{sroka2019} introduced a
    finite-difference based, open-source, block-structured MR-driven AMR code.
    Deiterding et~al.\ \cite{deiterding2020} showed the benefits of using MR
    indicators in the AMROC code for several multi-dimensional test problems.
    The trend of MR schemes towards either patch- or block-based AMR can be
    understood as preference for more efficient memory access patterns and
    scalable computing. However, the mesh efficiency issues often present with
    these AMR formats has yet to be sufficiently addressed.

    % punchline
    In the current work, we present a hybrid adaptive MR (HAMR) approach that
    combines the scalability of block-structured AMR with the potential
    efficiency increase offered by the solver adaptive approach.\footnote{The
    source code required to reproduce the results presented in this work is made
    available at
    \href{https://github.com/blg13/HAMR-FLASH}{https://github.com/blg13/HAMR-FLASH}}
    This scheme uses MR indicators not only to adapt the mesh, but also to
    identify solution regions  where otherwise costly computations involving
    various physics can be replaced safely with interpolation without
    sacrificing the overall solution accuracy. We apply the solver adaptive
    approach to the hydrodynamic fluxes as well as to \emph{composite functions}
    of the solution, in particular the reactive source term and the EoS.
    Although we demonstrate this approach in the context of compressible
    reactive turbulence, other physics solvers can be treated in the same way.
    We use several benchmark problems to evaluate computational cost savings.

    % description of paper
    The paper is organized as follows. In \cref{sec:preliminaries} we introduce
    the system of coupled, nonlinear PDEs that are the subject of our
    investigation, as well as the finite volume (FV) scheme used to numerically
    solve those equations. We also provide an overview of the basics of MR
    adaptive schemes. In \cref{sec:hybridmr}, we present the HAMR scheme, which
    uses MR smoothness indicators not only to adapt the mesh but also to reduce
    the computational cost of the physics solvers. We apply the new scheme to
    select problems in \cref{sec:results}, and in \cref{sec:discussion} we
    provide analysis in terms of solution accuracy and overall computational
    efficiency.
\section{Preliminaries}
\label{sec:preliminaries}
%
%
%
    % intro do overall computational approach
    We begin our presentation by defining the basic components of the new
    scheme, including the model PDEs, FV framework, and the MR background.

    \subsection{Reactive flow equations}

        % introduce conservation laws
        The present approach is concerned with numerically solving the Euler equations for
        compressible, reactive flows:
        \begin{subequations}\label{eqn:euler}
        \begin{align}
            \rho_{t} + \nabla \cdot (\rho \bm{v}) &= 0, \label{eqn:mass} \\
            (\rho \bm{v})_{t} + \nabla \cdot (\rho \bm{v}
                \bm{v}^{T} + p\bm{I}) &= \bm{0}, \label{eqn:momentum} \\
            (\rho \bm{X})_{t} + \nabla \cdot \left( \rho \bm{X}
            \bm{v} \right) &= \bm{R}, \label{eqn:species} \\
            (\rho E)_{t} + \nabla \cdot \left( \rho E +
                p \right) \bm{v} &= \dot{Q}. \label{eqn:energy}
        \end{align}
        \end{subequations}
        Here $\rho$ is the mass density of the gas mixture, $\bm{v}$ is the
        velocity vector, $p$ is the pressure, $E$ is the specific total energy,
        $\bm{X}$ is the vector of species mass fractions, $\bm{R}$ is the vector
        of species reaction rates, and $\dot{Q}$ is the energy source term.  The
        mass fraction for the $i^{\mathrm{th}}$ species can be written as $X_{i}
        = \rho_{i} / \rho$, where $\rho_{i}$ is the corresponding mass density,
        meaning that the species must collectively satisfy the constraint
        $\sum_{i=1}^{\mathcal{N}_{\mathrm{species}}} X_{i} = 1$, where
        $\mathcal{N}_{\mathrm{species}}$ is the number of species considered.

        The total specific energy is calculated as the sum of the specific
        internal energy and the kinetic energy as  $E = e + \frac{1}{2} \bm{v}
        \bm{v}^{T}$.  The above system of PDEs is closed using a suitable EoS
        which relates the pressure to density, internal energy, and composition
        (see below and \cref{sec:adaptivesources}).

    \subsection{Finite volume method}
    \label{sec:fvm}

        The equations \cref{eqn:euler} are discretized using the standard FV
        approach. Given a uniform discretization of the domain $x \in [x_{a},
        x_{b}]$ with cell size $h = (x_{b} - x_{a}) / N$, where $N$ is the number of
        cells, the following semi-discrete scheme is considered:
        \begin{equation}
            u_{i}^{n+1} = u_{i}^{n} - \frac{1}{h}
                \int_{t_{n}}^{t_{n+1}} \left(
                \hat{f}_{i} -
                \hat{f}_{i-1} \right) dt +
                \int_{t_{n}}^{t_{n+1}}
                \hat{s}_{i} d t.
            \label{eqn:fv_scheme}
        \end{equation}
        Here, $u_{i}^{n}$ is an approximation to the average of the exact
        solution, which we denote as $q(x,t)$, in the control volume $[x_{i-1},
        x_{i}]$ at $t_{n}$.  The numerical flux approximates the exact flux
        function as
        \begin{equation}
            \hat{f}_{i} \coloneqq \hat{f} \left( u_{i-k}, \dots,
                u_{i+k+1} \right) \approx f \left( q(x_{i},t) \right),
            \label{eqn:fluxes}
        \end{equation}
        where $2k$ is the number of cells comprising the reconstruction stencil.
        Likewise, the term
        \begin{equation}
            \hat{s}_{i} \coloneqq s(u_{i}) \approx
                \frac{1}{h} \int_{x_{i-1}}^{x_{i}} s(q(x,t))
                dx
            \label{eqn:source}
        \end{equation}
        is an approximate average of the exact source term within the control
        volume.  We present these equations in the semi-discrete form to
        emphasize that the fluxes and source terms can be handled separately in
        time using standard operator splitting approaches (see,
        \cite{macnamara+16}, and references therein).

    \subsection{Multiresolution decomposition}

        The MR representation has been used for the purpose of dynamic mesh
        adaptation in a number of FV-based approaches. These approaches rely on
        the local MR smoothness indicators to adapt the mesh. The smoothness
        indicators are obtained on a hierarchy of nested meshes. We outline the
        MR decomposition presented in \cite{harten1994}, which is appropriate
        for the current one-dimensional presentation, and refer the reader to
        \cite{bihari1997} and \cite{roussel2003} for the two- and
        three-dimensional extensions, respectively. The mesh hierarchy is
        defined as
        \begin{equation}
            \bm{\mathcal{G}}_{l} = \left\{ x_{l,i} \right\}_{i=0}^{N_{l}},
            \quad x_{l,i} = x_{a} + i \cdot h_{l},
            \quad h_{l} = 2^{(L-l)} h_{L},
            \quad N_{l} = N_{L} / 2^{(L-l)},
            \label{eqn:hierarchy}
        \end{equation}
        where $L$ is the finest level of mesh resolution allowed and $1 \leq l
        \leq L$. The objective in introducing \cref{eqn:hierarchy} is to
        represent the fine-grid data as a sum of data averages on the coarsest
        level plus a series of differences defined on finer levels. Given cell
        averages on the finest level of resolution, $\bm{u}^{n}_{L} =
        \left\{u_{L,i}^{n}\right\}_{i=1}^{N_{L}}$, where $n$ is the time index,
        the decomposition is performed using the following set of mappings
        beginning with level $l = L-1$ and ending with level $l = 1$.  First,
        cells on level $l+1$ are \emph{projected} onto the coarser grid level
        $l$ by means of volume weighted averaging. Then, approximate cell
        averages on the finer level $l+1$ are \emph{predicted} by an
        average-interpolating polynomial constructed with data on level $l$. In
        this work we use the third-order polynomial commonly found in the
        literature.

        The smoothness of the data is assessed using detail coefficients.  These
        coefficients are computed as the difference between actual and predicted
        values as
        \begin{equation}
            d_{l,i}^{n} =
                u_{l+1,2i}^{n} -
                \tilde{u}_{l+1,2i}^{n},
            \label{eqn:detail}
        \end{equation}
        where $\tilde{u}_{l+1,2i}$ is the prediction based on coarser data (see
        \cite{harten1994} for one-dimensional formulae). The detail coefficients
        are a measure of the local regularity of the solution, and their values
        decay in relation to that solution regularity as the mesh resolution
        increases \cite{harten1994}.

        % describe truncation procedure
        Compression of the MR representation is achieved by simply setting to
        zero the coefficients whose magnitudes are below a prescribed threshold.
        In this work we adopt the scale dependent threshold of Harten,
        $\varepsilon_{l} = \varepsilon / 2^{\left(L-l \right)}$, where
        $\varepsilon$ ($= \varepsilon_{L}$) is the prescribed tolerance.  The
        remaining non-zero detail coefficients form a set of \emph{significant}
        detail coefficients and encode the information necessary to approximate
        the fine-grid data to a prescribed level of accuracy.  This set of
        significant detail coefficients is given by $\bm{\mathcal{D}}^{n} =
        \left\{(l,i):\ |d^{n}_{l,i}| > \varepsilon_{l} \right\}$.

        Due to the time-dependent character of the solution, the set of cells
        determined to be significant at a given time will in general be
        inadequate for describing future states. In the case of the CFL-limited
        time step integration, the solution structure evolves slowly with a rate
        of at most one cell per step. Thus one can safely assume that the MR
        decomposition will correctly capture relevant solution structures at
        future times by anticipating the amount of variation in the neighborhood
        \cite{harten1994, cohen2003}.

        To maintain a reliable MR representation, the set $\bm{\mathcal{D}}^{n}$
        is extended to include coefficients that belong to the \textit{a~priori}
        unknown set $\bm{\mathcal{D}}^{n+1}$ as well. Thus a prediction set
        $\tilde{\bm{\mathcal{D}}}^{n+1}$ is created with the intent of
        satisfying the condition,
        \begin{equation}
            \bm{\mathcal{D}}^{n} \cup \bm{\mathcal{D}}^{n+1} \subset
            \tilde{\bm{\mathcal{D}}}^{n+1}.
            \label{eqn:mr_reliability}
        \end{equation}
        The set $\tilde{\bm{\mathcal{D}}}^{n+1}$ can be estimated by
        constructing a mask, $\bm{\mathcal{M}}^{n}$, initially set to
        $\bm{\mathcal{D}}^{n}$ and then expanded by adding a \emph{buffer
        region} such that
        \begin{equation}
          \bm{\mathcal{M}}^{n} \coloneqq \bm{\mathcal{D}}^{n} \cup
                \left\{ (l,i) :
                |i-\hat{\I}| \leq
                \mathcal{N}_{\mathrm{buffer}} \right\},
                \label{eqn:mask}
        \end{equation}
        where $\mathcal{N}_{\mathrm{buffer}}$ is the size of
        the buffer region and $(l,\hat{\I}) \in \bm{\mathcal{D}}^{n}$.  For
        added reliability of the MR scheme, more cells can be included in the
        buffer region to account for possible rapid development of the solution
        structure \cite{harten1994,cohen2003}.

        As alluded to earlier in this section, the buffer size in practical
        applications is determined by the numerical speed of signal propagation
        on the mesh. For HCLs,  this constraint is expressed by the usual CFL
        condition \cite{courant1967}. For that reason, if the MR decomposition
        is performed at every timestep, then the minimal buffer size required
        to satisfy the reliability condition \cref{eqn:mr_reliability} should be
        $\mathcal{N}_{\mathrm{buffer}} = 1$. If one wishes to refine less often,
        then the buffer size should be proportionally increased.
\section{Hybrid adaptive multiresolution scheme}
\label{sec:hybridmr}
%
%
%
        % show the adaptive MR FV scheme here
        In this section we introduce the proposed HAMR scheme, which combines
        the MR-driven block-structured AMR with the solver-adaptive ideas
        initially proposed by Harten \cite{harten1994}. To describe this
        approach we must consider the reference FV scheme \cref{eqn:fv_scheme}
        on the MR hierarchy \cref{eqn:hierarchy}. The MR FV scheme is defined on
        each level $l$ of the hierarchy as,
        \begin{subequations}
            \begin{align}
                u^{*}_{l,i} & = u^{n}_{l,i} - \frac{\Delta
                    t}{h_{l}} \left( \bar{f}_{l,i} -
                    \bar{f}_{l,i-1} \right),
                    \label{eqn:multiscale_fluxes} \\
                u^{n+1}_{l,i} & = u^{*}_{l,i} +
                    \Delta t \hat{s}^{n+1}_{l,i},
                    \label{eqn:multiscale_sources}
            \end{align}
        \end{subequations}
        where $i = 1, \dots, N_{l}$ and $\bar{f}_{l,i}$ is the time-averaged
        numerical flux through the respective interface. Here we utilize the
        Godunov operator splitting approach \cite{godunov1959}, with the
        reactive source terms being treated implicitly \footnote{While we
        demonstrate the HAMR approach using single stage schemes
        \cref{eqn:multiscale_fluxes} - \cref{eqn:multiscale_sources}, in
        practice, higher order multi-stage schemes may be used to integrate
        individual physics solvers in time.}.  The question of how to compute
        the fluxes and sources on the adaptive hierarchy is briefly reviewed. We
        summarize three main approaches available in the literature, and then
        describe how the new HAMR scheme operates.

        The first approach, pioneered by Harten \cite{harten1994} and which we
        refer to as solver adaptive, evolves the numerical solution on the
        finest mesh level but reduces computational effort in smooth regions by
        replacing direct flux calculations with interpolation from values
        previously computed on coarser levels. The fluxes on coarse levels are
        computed as
        \begin{equation}
            \bar{f}_{l,i} = \bar{f} \left(
                \tilde{u}_{L,2^{L-l}i-k+1}^{n}, \dots,
                \tilde{u}_{L,2^{L-l}i+k}^{n} \right).
                \label{eqn:coarse_flux}
        \end{equation}
        Here $2^{L-l} i$ is the index corresponding to interface $i$ on the
        coarse level, as represented on the finest level, $L$. The
        MR-interpolated solution, $\tilde{\bm{u}}_{L}^{n}$, is obtained via the
        inverse MR decomposition based on the significant set of cells
        $\bm{\mathcal{M}}^{n}$ (note that in this approach, the inverse MR
        decomposition needs only to be considered for analysis purposes, as the
        fine-scale solution data is always available).  In a closely related
        work involving reactive systems, Bihari \& Harten \cite{bihari1999} also
        evaluate source terms (explicitly) as a function of fine-scale data as
        \begin{equation}
            \hat{s}_{l,i}^{n} = \frac{1}{2^{L-l}}
                \sum_{\hat{\I}=1}^{2^{L-l}}
                \hat{s}_{L, 2^{L-l}(i-1) + \hat{\I}}^{n}.
                \label{eqn:coarse_source_exact}
        \end{equation}
        Naturally this approach has the inherent disadvantage of the
        computational complexity being dependent on the finest mesh level.

        The second type of approach, initially developed by Gottschlich-M\"uller
        \& M\"uller \cite{gottschlich-muller1999} and Cohen et~al.\
        \cite{cohen2003}, contrasts with the approach of Harten by introducing
        an adaptive, nonuniform discretization based on the MR indicators.
        In this approach, \cref{eqn:coarse_flux} is used in conjunction with the
        inverse MR decomposition to compute coarse-scale fluxes as a function of
        fine-scale data in order to avoid the discretization error associated
        with the local level of resolution.  This type of procedure is referred
        to as the \emph{exact local reconstruction}. A variation of this
        approach, explored by Hovhannisyan \& M\"uller \cite{hovhannisyan2010}
        and referred to as the \emph{approximate reconstruction strategy},
        utilizes an interpolating polynomial constructed from local averages in
        conjunction with an appropriate quadruate rule to reduce the
        computational complexity while maintaining accuracy.

        The third, simplest and most commonly used strategy of handling fluxes
        and source terms on the adaptive mesh is to provide the solvers with
        local solution averages. This \emph{direct evaluation} strategy
        is attractive because it has the lowest computational complexity.
        However, it is expected to incur the greatest overall error -- this is
        because the contributing discretization error is set by the local
        resolution (which might be coarser than the finest allowed resolution),
        while in the other two approaches described above, the error is always
        set by the finest allowed mesh resolution (by using interpolation
        procedures).

        The HAMR approach developed in this work adapts the block-structured
        mesh according to the MR mask \cref{eqn:mask}, and employs a hybrid
        strategy for the evaluation of fluxes and composite functions on each
        individual mesh block. Principally, the evolution of the solution on
        each mesh block is handled according to the direct evaluation strategy.
        However, computations are further accelerated locally by the use of the
        solver adaptive approach. In particular, the fluxes are computed in the
        same fashion as Harten's original approach, but in this case the flux
        formula \cref{eqn:coarse_flux} is applied with respect to the block's
        level, $l$, rather than the finest level, $L$.  Composite functions are
        treated similarly.

        Our description now turns to the construction of the adaptive mesh
        and the procedure for adaptively computing fluxes and composite
        functions on that mesh.

        \subsection{Block-structured adaptive mesh refinement}
        \label{sec:amrstructure}

            % describe the SAMR
            We consider dynamically adaptive block-structured mesh hierarchies.
            For such a mesh hierarchy, the natural data representation is a
            tree-like structure (binary tree, quadtree, or octree in one, two,
            or three dimensions, respectively).  The blocks in the hierarchy
            each consist of a fixed number of computational cells,
            $N_{\mathrm{c}}$ in each dimension, and the block refinement
            produces $2^d$ children blocks, where $d$ is the number of
            dimensions.  The numerical solution is obtained and evolved on
            blocks that are no further refined, referred to as \emph{leaf
            blocks}.  In addition to the computational cells, blocks must be
            equipped with \emph{ghost cells} to provide the PDE solvers with
            data beyond the block boundaries necessary for operations using long
            data stencils.

            \subsubsection{Local MR hierarchy}
            \label{sec:mrsubhierarchy}

                We decompose the global MR hierarchy (spanning the computational
                domain) into a collection of \emph{local MR hierarchies}, with
                each leaf block being assigned a partition of the global
                hierarchy in space and in scale. Each leaf block serves the dual
                purpose of handling a part of the solution and a respective part
                of the MR decomposition. On each leaf block there are
                $\mathcal{L}$ levels defined, with the finest level
                corresponding to the block's global MR (or AMR) level.

\begin{figure}
                    \center
                    \scalebox{0.85}{
                        \begin{tikzpicture}[thick,scale=0.35, every node/.style={scale=0.6}]

    % variables
    \def\xl{-8.0}
    \def\xr{8.0}
    \def\y{0.0}
    \def\yy{-3.0}
    \def\yyy{-6.0}
    \def\ts{0.5}
    \def\op{0.35}
    \def\fx{0.15}
    
    % draw ghost cells
    \draw [mygray] (\xl-4,\y+\ts/2) --(\xr+4,\y+\ts/2);
    \draw [mygray] (\xl-4,\yy+\ts/2) --(\xr+4,\yy+\ts/2);
    \draw [mygray] (\xl-4,\yyy+\ts/2) --(\xr+4,\yyy+\ts/2);
    \draw [mygray] (\xl-1.0,\y) --(\xl-1.0,\y+\ts);
    \draw [mygray] (\xr+1.0,\y) --(\xr+1.0,\y+\ts);
    \draw [mygray] (\xl-2.0,\y) --(\xl-2.0,\y+\ts);
    \draw [mygray] (\xr+2.0,\y) --(\xr+2.0,\y+\ts);
    \draw [mygray] (\xl-3.0,\y) --(\xl-3.0,\y+\ts);
    \draw [mygray] (\xr+3.0,\y) --(\xr+3.0,\y+\ts);
    \draw [mygray] (\xl-4.0,\y) --(\xl-4.0,\y+\ts);
    \draw [mygray] (\xr+4.0,\y) --(\xr+4.0,\y+\ts);

    % ghost cells at level l=1
    \draw [mygray] (\xl-2.0,\yy) --(\xl-2.0,\yy+\ts);
    \draw [mygray] (\xr+2.0,\yy) --(\xr+2.0,\yy+\ts);
    \draw [mygray] (\xl-4.0,\yy) --(\xl-4.0,\yy+\ts);
    \draw [mygray] (\xr+4.0,\yy) --(\xr+4.0,\yy+\ts);

    % level l=2
    \draw [mygray] (\xl-4.0,\yyy) --(\xl-4.0,\yyy+\ts);
    \draw [mygray] (\xr+4.0,\yyy) --(\xr+4.0,\yyy+\ts);

    % draw grids
    \draw (\xl,\y+\ts/2) --(\xr,\y+\ts/2);
    \draw (\xl,\yy+\ts/2) --(\xr,\yy+\ts/2);
    \draw (\xl,\yyy+\ts/2) --(\xr,\yyy+\ts/2);
    
    % draw cells for max level
    \draw (\xl,\y) --(\xl,\y+\ts);
    \draw (\xl+1.0,\y) --(\xl+1.0,\y+\ts);
    \draw (\xl+2.0,\y) --(\xl+2.0,\y+\ts);
    \draw (\xl+3.0,\y) --(\xl+3.0,\y+\ts);
    \draw (\xl+4.0,\y) --(\xl+4.0,\y+\ts);
    \draw (\xl+5.0,\y) --(\xl+5.0,\y+\ts);
    \draw (\xl+6.0,\y) --(\xl+6.0,\y+\ts);
    \draw (\xl+7.0,\y) --(\xl+7.0,\y+\ts);
    \draw (\xl+8.0,\y) --(\xl+8.0,\y+\ts);
    \draw (\xl+9.0,\y) --(\xl+9.0,\y+\ts);
    \draw (\xl+10.0,\y) --(\xl+10.0,\y+\ts);
    \draw (\xl+11.0,\y) --(\xl+11.0,\y+\ts);
    \draw (\xl+12.0,\y) --(\xl+12.0,\y+\ts);
    \draw (\xl+13.0,\y) --(\xl+13.0,\y+\ts);
    \draw (\xl+14.0,\y) --(\xl+14.0,\y+\ts);
    \draw (\xl+15.0,\y) --(\xl+15.0,\y+\ts);
    \draw (\xl+16.0,\y) --(\xl+16.0,\y+\ts);

    % lower level cells
    \draw (\xl,\yy) --(\xl,\yy+\ts);
    \draw (\xl+2.0,\yy) --(\xl+2.0,\yy+\ts);
    \draw (\xl+4.0,\yy) --(\xl+4.0,\yy+\ts);
    \draw (\xl+6.0,\yy) --(\xl+6.0,\yy+\ts);
    \draw (\xl+8.0,\yy) --(\xl+8.0,\yy+\ts);
    \draw (\xl+10.0,\yy) --(\xl+10.0,\yy+\ts);
    \draw (\xl+12.0,\yy) --(\xl+12.0,\yy+\ts);
    \draw (\xl+14.0,\yy) --(\xl+14.0,\yy+\ts);
    \draw (\xl+16.0,\yy) --(\xl+16.0,\yy+\ts);
    
    % lower level cells
    \draw (\xl,\yyy) --(\xl,\yyy+\ts);
    \draw (\xl+4.0,\yyy) --(\xl+4.0,\yyy+\ts);
    \draw (\xl+8.0,\yyy) --(\xl+8.0,\yyy+\ts);
    \draw (\xl+12.0,\yyy) --(\xl+12.0,\yyy+\ts);
    \draw (\xl+16.0,\yyy) --(\xl+16.0,\yyy+\ts);
 
    % arrows indicating flux interpolation dependency
    \draw[decoration={brace,mirror},decorate]
        (\xr-2,\y-1.0) -- node {}(\xr,\y-1.0);
    \draw[->,line width=0.2mm,>=stealth] (\xr-3,\yy+\ts) -- (\xr-1.3,\y-1.5);
    \draw[->,line width=0.2mm,>=stealth] (\xr-1,\yy+\ts) -- (\xr-1,\y-1.5);
    \draw[mygray,->,line width=0.2mm,>=stealth] (\xr+1,\yy+\ts) --
    (\xr-0.7,\y-1.5);

    \draw[decoration={brace,mirror},decorate]
        (\xr-4,\yy-1.0) -- node {}(\xr,\yy-1.0);
    \draw[->,line width=0.2mm,>=stealth] (\xr-6,\yyy+\ts) -- (\xr-2.3,\yy-1.5);
    \draw[->,line width=0.2mm,>=stealth] (\xr-2,\yyy+\ts) -- (\xr-2,\yy-1.5);
    \draw[mygray,->,line width=0.2mm,>=stealth] (\xr+2,\yyy+\ts) -- (\xr-1.7,\yy-1.5);

    \draw[decoration={brace,mirror},decorate]
        (\xl+0,\y-1.0) -- node {}(\xl+2,\y-1.0);
    \draw[mygray,->,line width=0.2mm,>=stealth] (\xl-1,\yy+\ts) -- (\xl+0.7,\y-1.5);
    \draw[->,line width=0.2mm,>=stealth] (\xl+1,\yy+\ts) -- (\xl+1,\y-1.5);
    \draw[->,line width=0.2mm,>=stealth] (\xl+3,\yy+\ts) -- (\xl+1.3,\y-1.5);

    \draw[decoration={brace,mirror},decorate]
        (\xl+0,\yy-1.0) -- node {}(\xl+4,\yy-1.0);
    \draw[mygray,->,line width=0.2mm,>=stealth] (\xl-2,\yyy+\ts) -- (\xl+1.6,\yy-1.5);
    \draw[->,line width=0.2mm,>=stealth] (\xl+2,\yyy+\ts) -- (\xl+2,\yy-1.5);
    \draw[->,line width=0.2mm,>=stealth] (\xl+6,\yyy+\ts) -- (\xl+2.4,\yy-1.5);

    % curly brace
    \draw[decoration={brace,raise=5pt},decorate]
        (\xr,\y+1.0) -- node[above=10pt] {\LARGE ghost region}(\xr+4.0,\y+1.0);
    \draw[decoration={brace,raise=5pt},decorate]
        (\xl-4.0,\y+1.0) -- node[above=10pt] {\LARGE ghost region}(\xl,\y+1.0);

    % nodes
    \node at (\xl-7.0,\y+0.25) {\Large $\mathcal{L}$};
    \node at (\xl-7.0,\yy+0.25) {\Large $\mathcal{L}-1$};
    \node at (\xl-7.0,\yyy+0.25) {\Large $\mathcal{L}-2$};

    % indices
    \node at (\xl+2.0,\yyy-0.5) {\Large $i=1$};
    \node at (\xl+6.0,\yyy-0.5) {\Large $2$};
    \node at (\xl+10.0,\yyy-0.5) {\Large $3$};
    \node at (\xl+14.0,\yyy-0.5) {\Large $4$};

    \node at (\xl+1.0,\yy-0.5) {\large $1$};
    \node at (\xl+3.0,\yy-0.5) {\large $2$};
    \node at (\xl+5.0,\yy-0.5) {\large $3$};
    \node at (\xl+7.0,\yy-0.5) {\large $4$};
    \node at (\xl+9.0,\yy-0.5) {\large $5$};
    \node at (\xl+11.0,\yy-0.5) {\large $6$};
    \node at (\xl+13.0,\yy-0.5) {\large $7$};
    \node at (\xl+15.0,\yy-0.5) {\large $8$};

    \node at (\xl+0.5,\y-0.5) {\large $1$};
    \node at (\xl+1.5,\y-0.5) {\large $2$};
    \node at (\xl+2.5,\y-0.5) {\large $3$};
    \node at (\xl+3.5,\y-0.5) {\large $4$};
    \node at (\xl+4.5,\y-0.5) {\large $5$};
    \node at (\xl+5.5,\y-0.5) {\large $6$};
    \node at (\xl+6.5,\y-0.5) {\large $7$};
    \node at (\xl+7.5,\y-0.5) {\large $8$};
    \node at (\xl+8.5,\y-0.5) {\large $9$};
    \node at (\xl+9.5,\y-0.5) {\large $10$};
    \node at (\xl+10.5,\y-0.5) {\large $11$};
    \node at (\xl+11.5,\y-0.5) {\large $12$};
    \node at (\xl+12.5,\y-0.5) {\large $13$};
    \node at (\xl+13.5,\y-0.5) {\large $14$};
    \node at (\xl+14.5,\y-0.5) {\large $15$};
    \node at (\xl+15.5,\y-0.5) {\large $16$};

\end{tikzpicture}}
                    \caption{The local MR hierarchy on the AMR leaf block, with
                    $N_{\mathrm{gc}} = 4$ ghost cells on each end, and the block
                    having $N_{\mathrm{c}} = 16$ cells.  Dependencies for the
                    one-dimensional, third-order prediction operator are shown near the boundaries,
                    illustrating the need for ghost cells (colored in gray), and the
                    limit on the number of levels that can be handled. At most three
                    levels can be handled in this case.}
                    \label{fig:block_decomp}
                \end{figure}
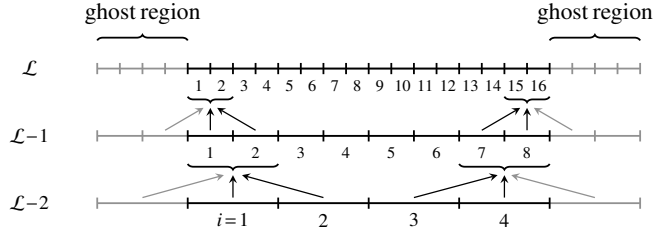

                The number of levels existing in the local MR hierarchy is
                limited by the number of cells in the block, $N_{\mathrm{c}}$
                ($= N_{\mathcal{L}}$) as well as the extent of the ghost region
                ($N_{\mathrm{gc}}$ cells in each direction). The latter
                limitation is due to the number of cells required by the
                prediction operator stencil.  As the local MR hierarchy depth
                increases, more cells are required from neighboring blocks to
                supply boundary data for the stencils.  This situation is
                illustrated in \cref{fig:block_decomp},
                where we show the local MR hierarchy for a given block, interpolation
                dependencies at the block's boundaries, and block ghost cell regions.
                Although the figure illustrates the situation in one dimension, the
                same dependencies carry over in the multi-dimensional case.

            \subsubsection{MR mask and block refinement and coarsening}
            \label{sec:flagging}

                The MR decomposition is computed on each local MR hierarchy as
                follows.  Let $\bm{\mathcal{B}}^{n}$ denote the set of AMR leaf
                block IDs at timestep $n$, and let $\bm{\mathcal{M}}^{n}_{b}$
                denote a component of the MR mask defined by \cref{eqn:mask}, where
                $b \in \bm{\mathcal{B}}^{n}$. One can define for each leaf block $b$
                a local-to-global mapping operator that maps the local hierarchy
                index space to the global hierarchy index space as
                $\mathcal{T}_{b} : \bm{\mathcal{M}}^{n}_{b} \mapsto
                \bm{\mathcal{M}}^{n}$.  The global MR mask can then be formed
                via the union of local masks as
                \begin{equation}
                    \bigcup_{b \in \bm{\mathcal{B}}^{n}}
                      \mathcal{T}_{b} \left( \bm{\mathcal{M}}^{n}_{b} \right)
                      \subset \bm{\mathcal{M}}^{n}.
                      \label{eqn:unionmasks}
                \end{equation}
                We note that while the MR decomposition performed on the local
                MR hierarchies produces a subset of the mask that would be
                obtained by performing the complete decomposition, the union of
                local masks is actually sufficient for the purpose of adapting
                the mesh. This is because the elements missing from the union of
                local masks correspond to cells belonging to inactive AMR
                (parent) blocks, and are not useful for mesh adaptation.

                Flagging for refinement or coarsening for block $b$ is done
                according to the following rules:
                \begin{enumerate}[leftmargin=6em]
                    \item[\emph{refine:}] if there exists any $(\mathcal{L}-1,i) \in
                        \bm{\mathcal{M}}_{b}^{n}$,
                    \item[\emph{coarsen:}] if there does not exist any $(\mathcal{L}-2,i)
                        \in \bm{\mathcal{M}}_{b}^{n}$.
                \end{enumerate}
                Note that we keep one finer mesh level than is otherwise indicated
                by the detail coefficients in order to capture any fine-scale
                information that might emerge during time integration.

        \subsection{Adaptive calculation of fluxes on the local MR hierarchy}
        \label{sec:adaptivefluxes}

            The computation of numerical fluxes on the local MR hierarchy
            follows the procedure outlined in \cite{harten1994}.  That is,
            the MR mask is used to identify interfaces where the flux may be
            accurately interpolated from values already obtained at
            interfaces corresponding to coarser grid levels.

            On the local MR hierarchy, the fluxes, when calculated directly, are
            calculated according to \cref{eqn:coarse_flux} but with level $L$
            being replaced by $\mathcal{L}$. This is consistent with the direct
            evaluation strategy discussed in \cref{sec:hybridmr}. Further
            savings are introduced by interpolating fluxes at odd-indexed
            interfaces according to the MR mask. Flux values at level $l+1$ may
            be interpolated in a pointwise sense from values obtained on level
            $l$ using
            \begin{equation}
                \bar{f}_{l+1,2i-1} \approx I_{r} (l,i; \bm{\bar{f}}_{l}),
                \label{eqn:flux_interp}
            \end{equation}
            where
            \begin{subequations}
            \begin{align}
                I_{r=0} (l,i; \bm{\bar{f}}_{l}) &\coloneqq \frac{1}{2} ( \bar{f}_{l,i-1} + \bar{f}_{l,i}), \quad i = 1, \dots, N_{l}, \label{eqn:lowordinterp} \\
                I_{r=1} (l,i; \bm{\bar{f}}_{l}) &\coloneqq
                \begin{cases}
                \frac{1}{16}
                    ( 5 \bar{f}_{l,i-1} + 15 \bar{f}_{l,i} - 5
                    \bar{f}_{l,i+1} + \bar{f}_{l,i+2}),& i = 1, \\
                \frac{1}{16}
                    ( -\bar{f}_{l,i-2} + 9 \bar{f}_{l,i-1} + 9
                    \bar{f}_{l,i} - \bar{f}_{l,i+1}),& i = 2, \dots,
                    N_{l}-1, \\
                \frac{1}{16}
                    ( \bar{f}_{l,i-3} - 5 \bar{f}_{l,i-2} + 15
                    \bar{f}_{l,i-1} + 5 \bar{f}_{l,i}),& i = N_{l} \label{eqn:hiwordinterp}.
                \end{cases}
            \end{align}
            \end{subequations}
            Here, the order of accuracy of the interpolation is $2r+2$. Note
            that near block boundaries, the stencil for the flux interpolation
            is modified so that fluxes beyond the block \footnote{In parallel
            computations, those fluxes might be computed by a different
            process.} are not required. In this case, the stencil is biased to
            include additional flux values within the block's domain.

            % describe the actual algorithm
            The calculation of fluxes on the local MR hierarchy is described in
            \cref{alg:fluxes}.
            \begin{algorithm}
                \begin{algorithmic}
                \FOR {$i = 0$ \text{to} $N_{1}$}
                    \STATE {$\bar{f}_{1,i} = \bar{f} (
                        u_{\mathcal{L},2^{\mathcal{L}-1}i-k+1}^{n}, \dots,
                        u_{\mathcal{L},2^{\mathcal{L}-1}i+k}^{n} )$}
                \ENDFOR
                \FOR {$l = 1$ \text{to} $\mathcal{L}-1$}
                    \FOR {$i = 1$ \text{to} $N_{l}$}
                        \IF {$(l,i) \in \bm{\mathcal{M}}_{b}^{n}$}
                            \STATE {$\bar{f}_{l+1,2i-1} = \bar{f}(
                                u_{\mathcal{L},2^{\mathcal{L}-(l+1)}(2i-1)-k+1}^{n}, \dots,
                                u_{\mathcal{L},2^{\mathcal{L}-(l+1)}(2i-1)i+k}^{n} )$}
                        \ELSE
                            \STATE {$\bar{f}_{l+1,2i-1} \approx I_{r}(l,i;
                                \bm{\bar{f}}_{l})$}
                        \ENDIF
                    \ENDFOR
                    \FOR {$i = 0$ \text{to} $N_{l}$}
                        \STATE {$\bar{f}_{l+1,2i} = \bar{f}_{l,i}$}
                    \ENDFOR
                \ENDFOR
                \end{algorithmic}
                \caption{The solver adaptive procedure applied to the
                    calculation of numerical fluxes on the local MR hierarchy
                    associated with block $b$.}
                \label{alg:fluxes}
            \end{algorithm}
            First, the fluxes corresponding to the coarsest level are
            calculated. This calculation is the result of the first loop.  Then,
            the scheme considers fluxes on the next finer level. In particular,
            flux values at interfaces corresponding to the odd indices are
            interpolated if the parent cell on the current level belongs to the
            mask. Otherwise, the fluxes are computed directly according to
            \cref{eqn:coarse_flux} (again, with level $L$ being replaced by
            $\mathcal{L}$). Even-index fluxes are copied from the coarser level.

        \subsection{Adaptive calculation of composite functions: source terms and EoS}
        \label{sec:adaptivesources}

            Next we consider an adaptive strategy for the evolution of source
            terms via \cref{eqn:multiscale_sources}. The source terms being
            considered are stiff, requiring implicit time integration, and in
            effect, iterative methods. This fact makes the design of an adaptive
            scheme which preserves the discretization error of the finest level
            quite difficult due to the need to update fine-scale averages during
            the iterative process. To avoid this complication, we use the direct
            evaluation strategy on coarser levels of the local MR hierarchy as
            well, evaluating the source term as a function of local averages as
            \begin{equation}
                \hat{s}_{l,i} \approx s(u_{l,i}).
                \label{eqn:coarse_source_naive}
            \end{equation}
            Solution updates \cref{eqn:multiscale_sources} are then performed
            according to the MR mask.  Finally, similarly to the hydrodynamic
            fluxes, required solution outputs on the finest level of the local MR
            hierarchy are interpolated from coarser levels using
            \begin{equation}
                u_{l+1,2i+\mu}^{n+1} \approx \bar{I}_{r} (l,i,\mu;
                    \bm{u}^{n+1}_{l}),
                \label{eqn:cellavg_interpolant}
            \end{equation}
            where
            \begin{align}
                \bar{I}_{r=1} (l,i,\mu; \bm{u}_{l}) &\coloneqq
                \begin{cases}
                    u_{l,i} - (-1)^{\mu} \left( \frac{3}{8} u_{l,i} - \frac{1}{2} u_{l,i+1}
                    + \frac{1}{8} u_{l,i+2} \right),& i = 1, \\
                    u_{l,i} + (-1)^{\mu} \frac{1}{8} \left( u_{l,i+1} - u_{l,i-1} \right),& i = 2, \dots,
                    N_{l}-1, \\
                    u_{l,i} + (-1)^{\mu} \left( \frac{1}{8} u_{l,i-2} - \frac{1}{2} u_{l,i-1}
                    + \frac{3}{8} u_{l,i} \right),& i = N_{l},
                \end{cases}
            \end{align}
            for $\mu \in \{-1,0\}$.  Because the interpolation of source term
            outputs uses a stencil of data from coarser MR levels, special
            treatment is needed near the block boundaries.  To avoid
            communication of solution updates in those regions, we again make
            use of biased stencils for the MR interpolation
            \footnote{Multi-dimensional biased average interpolating formulae
            are available upon request.}.

            In general, the EoS routine iteratively solves a nonlinear equation
            that expresses pressure and temperature as a function of
            hydrodynamic state. The solver adaptive MR scheme for the EoS works
            on the local MR hierarchy in essentially the same way as the source
            terms, with EoS outputs being interpolated from coarser ones in
            smooth regions using \cref{eqn:cellavg_interpolant}.

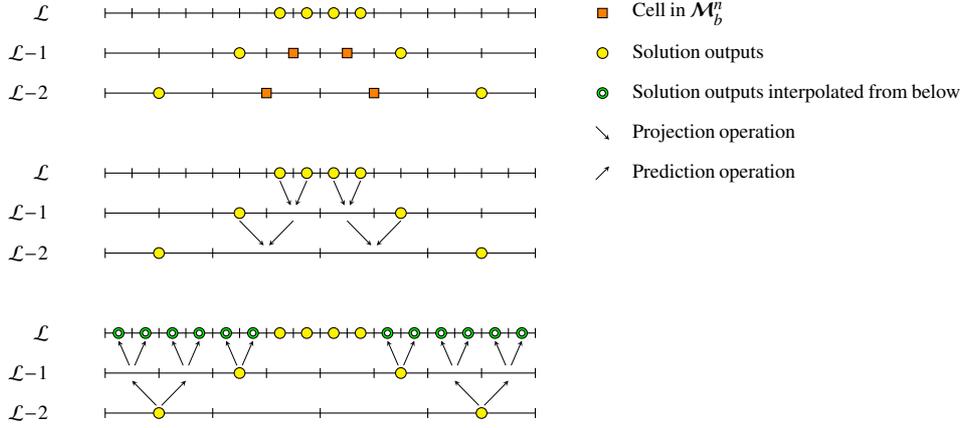
\begin{figure}
                \center
                \resizebox{0.77\textwidth}{!}{
                    \begin{tikzpicture}[thick,scale=0.35, every node/.style={scale=0.6}]

    % variables
    \def\xl{-16.0}
    \def\xr{16.0}
    \def\dx{2}
    \def\dxx{4}
    \def\dxxx{8}
    \def\y{0.0}
    \def\yy{-3.0}
    \def\yyy{-6.0}
    \def\ts{0.75}
    \def\op{0.35}
    \def\fx{0.15}
    \def\ra{0.40}
    \def\sq{0.35}
    
    % draw grid axes
    \draw (\xl,\y+\ts/2) --(\xr,\y+\ts/2);
    \draw (\xl,\yy+\ts/2) --(\xr,\yy+\ts/2);
    \draw (\xl,\yyy+\ts/2) --(\xr,\yyy+\ts/2);

    % draw cells for max level
    \foreach \b in {0,...,16}{
        \draw (\xl+\b*\dx,\y) --(\xl+\b*\dx,\y+\ts);
    }
    
    % lower level cells
    \foreach \b in {0,...,8}{
        \draw (\xl+\b*\dxx,\yy) --(\xl+\b*\dxx,\yy+\ts);
    }
    
    % even lower level cells
    \foreach \b in {0,...,4}{
        \draw (\xl+\b*\dxxx,\yyy) --(\xl+\b*\dxxx,\yyy+\ts);
    }

    % draw dots where source terms are computed
    \draw [fill=yellow] (\xl+6.5*\dx,\y+\ts/2) circle [radius=\ra];
    \draw [fill=yellow] (\xl+7.5*\dx,\y+\ts/2) circle [radius=\ra];
    \draw [fill=yellow] (\xl+8.5*\dx,\y+\ts/2) circle [radius=\ra];
    \draw [fill=yellow] (\xl+9.5*\dx,\y+\ts/2) circle [radius=\ra];
    \draw [fill=yellow] (\xl+5*\dx,\yy+\ts/2) circle [radius=\ra];
    \draw [fill=yellow] (\xl+11*\dx,\yy+\ts/2) circle [radius=\ra];
    \draw [fill=orange] (\xl+7*\dx-\sq,\yy+\ts/2-\sq) rectangle
        (\xl+7*\dx+\sq,\yy+\ts/2+\sq);
    \draw [fill=orange] (\xl+9*\dx-\sq,\yy+\ts/2-\sq) rectangle
        (\xl+9*\dx+\sq,\yy+\ts/2+\sq);
    \draw [fill=yellow] (\xl+2*\dx,\yyy+\ts/2) circle [radius=\ra];
    \draw [fill=yellow] (\xl+14*\dx,\yyy+\ts/2) circle [radius=\ra];
    \draw [fill=orange] (\xl+6*\dx-\sq,\yyy+\ts/2-\sq) rectangle
        (\xl+6*\dx+\sq,\yyy+\ts/2+\sq);
    \draw [fill=orange] (\xl+10*\dx-\sq,\yyy+\ts/2-\sq) rectangle
        (\xl+10*\dx+\sq,\yyy+\ts/2+\sq);

    % legend
    \draw [fill=orange] (\xr+5.0-\sq,\y+\ts/2-\sq) rectangle
        (\xr+5.0+\sq,\y+\ts/2+\sq);
    \draw[fill=yellow] (\xr+5.0,\y+\ts/2-3) circle [radius=\ra];
    \draw[fill=green] (\xr+5.0,\y+\ts/2-6) circle [radius=\ra];
    \draw[fill=white] (\xr+5.0,\y+\ts/2-6) circle [radius=0.22];
    \draw[->,line width=0.25mm,>=stealth] (\xr+4.5,\y+\ts/2-8.5)
        --(\xr+5.5,\y+\ts/2-9.5);
    \draw[->,line width=0.25mm,>=stealth] (\xr+4.5,\y+\ts/2-12.5)
        --(\xr+5.5,\y+\ts/2-11.5);
    \node[anchor=west] at (\xr+7,\y+\ts/2) {\Huge Cell in
            $\bm{\mathcal{M}}_{b}^{n}$};
    \node[anchor=west] at (\xr+7,\y+\ts/2-3) {\Huge Solution outputs};
    \node[anchor=west] at (\xr+7,\y+\ts/2-6) {\Huge Solution outputs interpolated from below};
    \node[anchor=west] at (\xr+7,\y+\ts/2-9) {\Huge Projection operation};
    \node[anchor=west] at (\xr+7,\y+\ts/2-12) {\Huge Prediction operation};

    % grid level labels
    \node[anchor=east] at (\xl-4,\y+\ts/2) {\Huge $\mathcal{L}$};
    \node[anchor=east] at (\xl-4,\yy+\ts/2) {\Huge $\mathcal{L}-1$};
    \node[anchor=east] at (\xl-4,\yyy+\ts/2) {\Huge $\mathcal{L}-2$};

    % draw block 2 grid
    \def\y{-12.0}
    \def\yy{-15.0}
    \def\yyy{-18.0}
    \draw (\xl,\y+\ts/2) --(\xr,\y+\ts/2);
    \draw (\xl,\yy+\ts/2) --(\xr,\yy+\ts/2);
    \draw (\xl,\yyy+\ts/2) --(\xr,\yyy+\ts/2);

    % draw cells for max level
    \foreach \b in {0,...,16}{
        \draw (\xl+\b*\dx,\y) --(\xl+\b*\dx,\y+\ts);
    }
    
    % lower level cells
    \foreach \b in {0,...,8}{
        \draw (\xl+\b*\dxx,\yy) --(\xl+\b*\dxx,\yy+\ts);
    }
    
    % even lower level cells
    \foreach \b in {0,...,4}{
        \draw (\xl+\b*\dxxx,\yyy) --(\xl+\b*\dxxx,\yyy+\ts);
    }

    % draw dots where source terms are computed
    \draw [fill=yellow] (\xl+6.5*\dx,\y+\ts/2) circle [radius=\ra];
    \draw [fill=yellow] (\xl+7.5*\dx,\y+\ts/2) circle [radius=\ra];
    \draw [fill=yellow] (\xl+8.5*\dx,\y+\ts/2) circle [radius=\ra];
    \draw [fill=yellow] (\xl+9.5*\dx,\y+\ts/2) circle [radius=\ra];
    \draw [fill=yellow] (\xl+5*\dx,\yy+\ts/2) circle [radius=\ra];
    \draw [fill=yellow] (\xl+11*\dx,\yy+\ts/2) circle [radius=\ra];
    \draw [fill=yellow] (\xl+2*\dx,\yyy+\ts/2) circle [radius=\ra];
    \draw [fill=yellow] (\xl+14*\dx,\yyy+\ts/2) circle [radius=\ra];
 
    % draw arrows
    \draw[->,line width=0.25mm,>=stealth] (\xl+7.5*\dx,\y-0.2)
        --(\xl+7.1*\dx,\yy+\ts+0.2);
    \draw[->,line width=0.25mm,>=stealth] (\xl+6.5*\dx,\y-0.2)
        --(\xl+6.9*\dx,\yy+\ts+0.2);
    \draw[->,line width=0.25mm,>=stealth] (\xl+9.5*\dx,\y-0.2)
        --(\xl+9.1*\dx,\yy+\ts+0.2);
    \draw[->,line width=0.25mm,>=stealth] (\xl+8.5*\dx,\y-0.2)
        --(\xl+8.9*\dx,\yy+\ts+0.2);
    \draw[->,line width=0.25mm,>=stealth] (\xl+7*\dx,\yy-0.2)
        --(\xl+6.1*\dx,\yyy+\ts+0.2);
    \draw[->,line width=0.25mm,>=stealth] (\xl+5*\dx,\yy-0.2)
        --(\xl+5.9*\dx,\yyy+\ts+0.2);
    \draw[->,line width=0.25mm,>=stealth] (\xl+11*\dx,\yy-0.2)
        --(\xl+10.1*\dx,\yyy+\ts+0.2);
    \draw[->,line width=0.25mm,>=stealth] (\xl+9*\dx,\yy-0.2)
        --(\xl+9.9*\dx,\yyy+\ts+0.2);
    
    % grid level labels
    \node[anchor=east] at (\xl-4,\y+\ts/2) {\Huge $\mathcal{L}$};
    \node[anchor=east] at (\xl-4,\yy+\ts/2) {\Huge $\mathcal{L}-1$};
    \node[anchor=east] at (\xl-4,\yyy+\ts/2) {\Huge $\mathcal{L}-2$};

    % draw block 3 grid
    \def\y{-24.0}
    \def\yy{-27.0}
    \def\yyy{-30.0}
    \draw (\xl,\y+\ts/2) --(\xr,\y+\ts/2);
    \draw (\xl,\yy+\ts/2) --(\xr,\yy+\ts/2);
    \draw (\xl,\yyy+\ts/2) --(\xr,\yyy+\ts/2);

    % draw cells for max level
    \foreach \b in {0,...,16}{
        \draw (\xl+\b*\dx,\y) --(\xl+\b*\dx,\y+\ts);
    }
    
    % lower level cells
    \foreach \b in {0,...,8}{
        \draw (\xl+\b*\dxx,\yy) --(\xl+\b*\dxx,\yy+\ts);
    }
    
    % even lower level cells
    \foreach \b in {0,...,4}{
        \draw (\xl+\b*\dxxx,\yyy) --(\xl+\b*\dxxx,\yyy+\ts);
    }

    % draw dots where source terms are computed
    \draw [fill=yellow] (\xl+2.0*\dx,\yyy+\ts/2) circle [radius=\ra];
    \draw [fill=yellow] (\xl+14.0*\dx,\yyy+\ts/2) circle [radius=\ra];
    \draw [fill=yellow] (\xl+5.0*\dx,\yy+\ts/2) circle [radius=\ra];
    \draw [fill=yellow] (\xl+11.0*\dx,\yy+\ts/2) circle [radius=\ra];
    \draw [fill=yellow] (\xl+6.5*\dx,\y+\ts/2) circle [radius=\ra];
    \draw [fill=yellow] (\xl+7.5*\dx,\y+\ts/2) circle [radius=\ra];
    \draw [fill=yellow] (\xl+8.5*\dx,\y+\ts/2) circle [radius=\ra];
    \draw [fill=yellow] (\xl+9.5*\dx,\y+\ts/2) circle [radius=\ra];

    \foreach \b in {0.5,1.5,2.5,3.5,4.5,5.5,10.5,11.5,12.5,13.5,14.5,15.5}{
        \draw [fill=green] (\xl+\b*\dx,\y+\ts/2) circle [radius=\ra];
        \draw [fill=white] (\xl+\b*\dx,\y+\ts/2) circle [radius=0.22];
    }

     % draw arrows
    \draw[->,line width=0.25mm,>=stealth] (\xl+2.1*\dx,\yyy+\ts+0.2)
    --(\xl+3*\dx,\yy-0.2);
    \draw[->,line width=0.25mm,>=stealth] (\xl+1.9*\dx,\yyy+\ts+0.2)
    --(\xl+1*\dx,\yy-0.2);
    \draw[->,line width=0.25mm,>=stealth] (\xl+14.1*\dx,\yyy+\ts+0.2)
    --(\xl+15*\dx,\yy-0.2);
    \draw[->,line width=0.25mm,>=stealth] (\xl+13.9*\dx,\yyy+\ts+0.2)
    --(\xl+13*\dx,\yy-0.2);
    \draw[->,line width=0.25mm,>=stealth] (\xl+1.1*\dx,\yy+\ts+0.2)
    --(\xl+1.5*\dx,\y-0.2);
    \draw[->,line width=0.25mm,>=stealth] (\xl+0.9*\dx,\yy+\ts+0.2)
    --(\xl+0.5*\dx,\y-0.2);
    \draw[->,line width=0.25mm,>=stealth] (\xl+3.1*\dx,\yy+\ts+0.2)
    --(\xl+3.5*\dx,\y-0.2);
    \draw[->,line width=0.25mm,>=stealth] (\xl+2.9*\dx,\yy+\ts+0.2)
    --(\xl+2.5*\dx,\y-0.2);
    \draw[->,line width=0.25mm,>=stealth] (\xl+5.1*\dx,\yy+\ts+0.2)
    --(\xl+5.5*\dx,\y-0.2);
    \draw[->,line width=0.25mm,>=stealth] (\xl+4.9*\dx,\yy+\ts+0.2)
    --(\xl+4.5*\dx,\y-0.2);
    \draw[->,line width=0.25mm,>=stealth] (\xl+13.1*\dx,\yy+\ts+0.2)
    --(\xl+13.5*\dx,\y-0.2);
    \draw[->,line width=0.25mm,>=stealth] (\xl+12.9*\dx,\yy+\ts+0.2)
    --(\xl+12.5*\dx,\y-0.2);
    \draw[->,line width=0.25mm,>=stealth] (\xl+15.1*\dx,\yy+\ts+0.2)
    --(\xl+15.5*\dx,\y-0.2);
    \draw[->,line width=0.25mm,>=stealth] (\xl+14.9*\dx,\yy+\ts+0.2)
    --(\xl+14.5*\dx,\y-0.2);
    \draw[->,line width=0.25mm,>=stealth] (\xl+11.1*\dx,\yy+\ts+0.2)
    --(\xl+11.5*\dx,\y-0.2);
    \draw[->,line width=0.25mm,>=stealth] (\xl+10.9*\dx,\yy+\ts+0.2)
    --(\xl+10.5*\dx,\y-0.2);

    % grid level labels
    \node[anchor=east] at (\xl-4,\y+\ts/2) {\Huge $\mathcal{L}$};
    \node[anchor=east] at (\xl-4,\yy+\ts/2) {\Huge $\mathcal{L}-1$};
    \node[anchor=east] at (\xl-4,\yyy+\ts/2) {\Huge $\mathcal{L}-2$};

\end{tikzpicture}}
                    \caption{The solver adaptive algorithm applied to the
                    adaptive calculation of composite functions. The operations
                    of the algorithm are illustrated for the local MR
                    hierarchy with three levels. The first step of the
                    algorithm (top panel) computes solution outputs at
                    cells (yellow disks) deemed necessary by the solver
                    adaptive mask (orange squares). The solution outputs are
                    then coarsened (middle panel) so that the data
                    necessary for the interpolation procedure is available.
                    Finally, the missing outputs (green rings) on the finest
                    level of the local MR hierarchy are interpolated (bottom
                    panel).}
                \label{fig:source_reconstruction}
            \end{figure}

            The procedure for the adaptive calculation of composite functions
            described above is illustrated in \cref{fig:source_reconstruction}.
            To summarize, the calculation of composite functions on the local MR
            hierarchy consists of the following steps:
            \begin{enumerate}[leftmargin=1.5cm]
                \item For each cell not in the mask, the composite function is
                    calculated directly if either the parent cell is
                    in the mask, or the cell is on the coarsest level (cf.
                    filled yellow
                    circles in \cref{fig:source_reconstruction}).
                \item Newly calculated outputs are projected to
                    the coarsest level (see middle panel
                    in \cref{fig:source_reconstruction}).
                \item Missing composite function outputs on the finest level are
                    interpolated from coarser levels (see bottom panel of
                    \cref{fig:source_reconstruction}).
            \end{enumerate}
            Once this procedure is complete, all solution outputs are available
            at the finest level of the local MR hierarchy.

        \subsection{HAMR error budget and its control}

            The error in the AMR solution is a sum of discretization error,
            truncation error, and perturbation error due to the interpolation of
            the solution from coarse to fine levels.  Similarly, the error in
            the HAMR solution is the sum of the aforementioned errors plus
            additional perturbation error due to interpolation of solver
            quantities (e.g.  fluxes), as well as additional discretization
            error due to the evaluation of composite functions at coarser levels
            of resolution. The integration of the solver adaptive component into
            the AMR scheme necessitates control over the additional error
            introduced.

            Given the averages $\bar{\bm{q}}_{l}^{n}$ of the exact solution on a
            given level $l$ at timestep $n$, and the averages $\bm{u}_{l}^{n}$
            of the HAMR solution, we define the solution error as
            \begin{equation}
                \bm{e}_{l}^{n} = \bar{\bm{q}}_{l}^{n}
                    - \bm{u}_{l}^{n}.
            \end{equation}
            Then the overall error budget may be written for the finest level,
            $\mathcal{L}$, of the local MR hierarchy as
            \begin{equation}
                \norm{\bm{e}_{\mathcal{L}}^{n+1}} \leq
                    \norm{\bm{e}_{\mathcal{L}}^{n}}
                    + g(\tau, C \varepsilon, \delta),
                \label{eqn:error_bound}
            \end{equation}
            where $g$ is generally a nonlinear function of the contributing
            error components. The three error components considered here are the
            bound on the local truncation error (LTE), $\tau$, the perturbation
            error due to MR interpolation, $C \varepsilon$, and the additional
            discretization error incurred by evaluating composite functions
            according to the direct evaluation strategy (as described in
            \cref{sec:hybridmr}), $\delta$. The LTE is a function of mesh
            resolution and the timestep size. In
            the perturbation error term, the constant $C$ is independent of the
            MR level (for more details, see Section 7 of \cite{harten1995}). We
            note that the additional discretization error term, $\delta$, also
            depends on $\varepsilon$ as the evaluation of the composite
            functions (see \cref{sec:adaptivesources}) on coarser levels only
            occurs in regions identified as smooth by the tolerance.

            The perturbation error and additional discretization error can be
            controlled independently by introducing a separate tolerance,
            \begin{equation}
                \tilde{\varepsilon} = \kappa \varepsilon,
            \end{equation}
            used solely for the solver adaptive scheme. Here $\kappa$ is a
            user-determined \emph{safety factor}. Given that the perturbation
            error term in \cref{eqn:error_bound} is now bound by $C
            \tilde{\varepsilon}$, it is clear that the appropriate choice of
            $\kappa$ depends inversely on $C$. If estimates of $\delta$, the
            coefficient $C$, and $\tau$ (when source terms and/or complicated
            equations of state are present) are provided, either through
            numerical testing or analysis, an upper bound on the safety factor
            may be obtained.  Because our goal is to demonstrate the viability
            of the HAMR approach for a wide range of problems and solvers, we
            examine the effect of systematically varying the safety factor on
            solution accuracy. We leave the \textit{a~priori} determination of
            an optimal safety factor for particular problems and solvers to a
            future study.
\section{Numerical experiments}
\label{sec:results}
%
%
%
        % preamble
        In order to demonstrate the characteristics of the newly proposed
        scheme, we consider a number of increasingly complex test problems. We
        begin with a one-dimensional pure hydrodynamics test problem and build
        upon this by including additional physics. The particular solvers chosen
        for these problems are discussed briefly.

        Given the discontinuous nature of the present problem set, we use the
        direct Eulerian variant of the piecewise parabolic method (PPM;
        \cite{colella1984}), as implemented in the \FLASH\ code
        \cite{fryxell2000} for computing the numerical fluxes. As we will refer
        to in later discussions of computational efficiency, the implementation
        in \FLASH\ consists of three main routines: \texttt{intrfc},
        \texttt{states}, and \texttt{rieman}. These routines implement the
        reconstruction of discrete solution data, calculation of effective left
        and right states, and the Riemann solver, respectively. For the reactive
        problems, we employ the $7$ isotope reaction network of
        \cite{timmes2000c}, and the thermodynamic properties of the plasma are
        described using the Helmholtz EoS \cite{timmes2000a}.

        \subsection{Interaction of two blast waves}
        \label{sec:blast2}

            The first application is the classic interacting blast waves problem
            of Woodward \& Colella \cite{woodward1984}. In this problem, the
            computational domain is divided into three sections of gas of
            constant density at rest, differing in pressure. Strong shock waves
            are generated across the interfaces dividing the two external
            sections.  These waves collide and produce a highly complex region
            of doubly-shocked gases that are separated by a newly formed contact
            discontinuity. The challenge in this case is to correctly capture
            the interaction region, and in particular the new material
            interface.

\begin{figure}
                \center
                \includegraphics[width=0.95\textwidth]{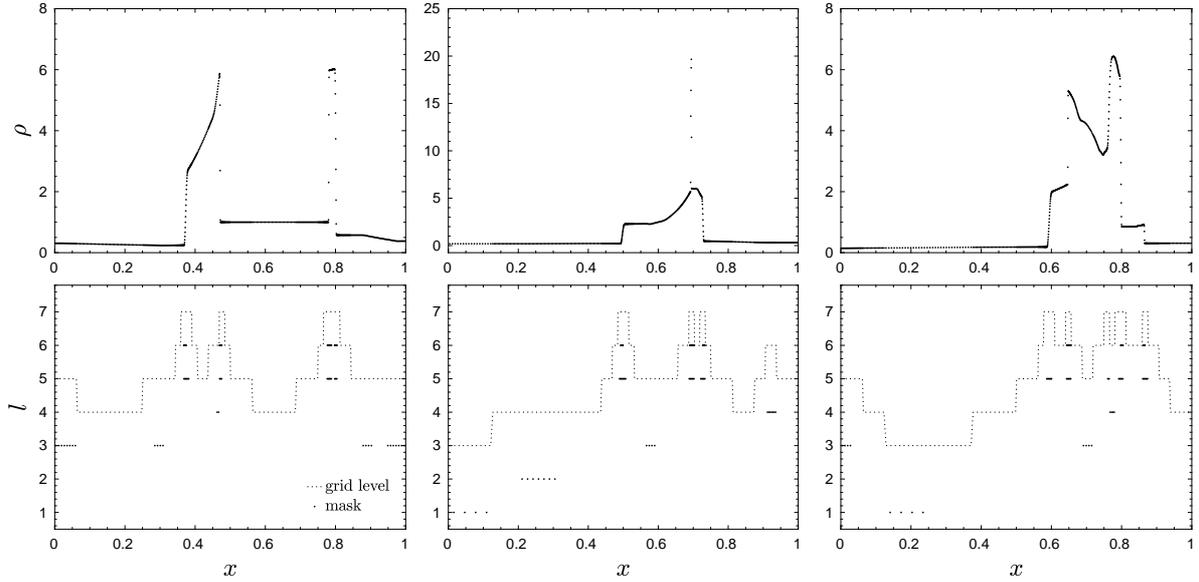}
                \caption{The HAMR solution for the two interacting blast waves
                    problem is shown for density, along with the mesh refinement
                    levels and MR mask (two levels per block), in top and bottom
                    rows, respectively. The evolution is shown before the shocks
                    collide ($t=0.016$; left column), at the time of the
                    collision ($t=0.028$; middle column), and at the final time
                    ($t=0.038$; right column).  The MR tolerance in this model
                    is $\varepsilon = 1 \times 10^{-2}$ and the safety factor is
                    $\kappa = 1 \times 10^{-1}$ . Note the sparse structure in
                    the MR mask, indicating that the solver adaptive approach
                    can be used to increase efficiency.}
                \label{fig:blast2}
            \end{figure}

            The domain is resolved using blocks containing $16$ mesh cells, with
            the base level covered by a single block. The effective resolution
            in our model is $N_{L} = 2048$. The mesh is adapted based on
            density, velocity, and pressure, and we use a MR tolerance of
            $\varepsilon = 1 \times 10^{-2}$, and a safety factor of $\kappa = 1
            \times 10^{-1}$ for solver adaptivity. In \cref{fig:blast2},
            the density profile (top row), as well as the mesh structure and
            associated MR mask (bottom row), is shown at early time ($t =
            0.016$\,s; left column panels) when the two blast waves are fully
            developed, at the time of their collision ($t=0.028$\,s; middle
            column panels), and at the final time ($t = 0.038$\,s; right column
            panels). At the end of simulation the flow structure is refined to
            the finest level across the interaction region, $x \in \left[0.6,
            0.8\right]$, bounded by the left contact discontinuity and
            right-moving transmitted shock.  At the adopted level of MR
            tolerance, the mesh resolves most of the perturbed sections of the
            flow to the finest level of resolution.

        \subsection{Hawley-Zabusky problem}
        \label{sec:hz}

            In the Hawley-Zabusky (HZ) problem \cite{hawley1989}, a planar shock
            wave obliquely strikes a material interface. In the "fast-slow" case
            of the HZ problem considered here, the shock is initially positioned
            in the low-density (high sound speed) material. As the time
            progresses, the shock gradually passes through the interface and
            refracts. This interaction results in the deposition of vorticity
            (driven by the baroclinic term), and in consequence material mixing,
            at the interface.

\begin{figure}
                \centering
                \includegraphics[width=0.96\textwidth]{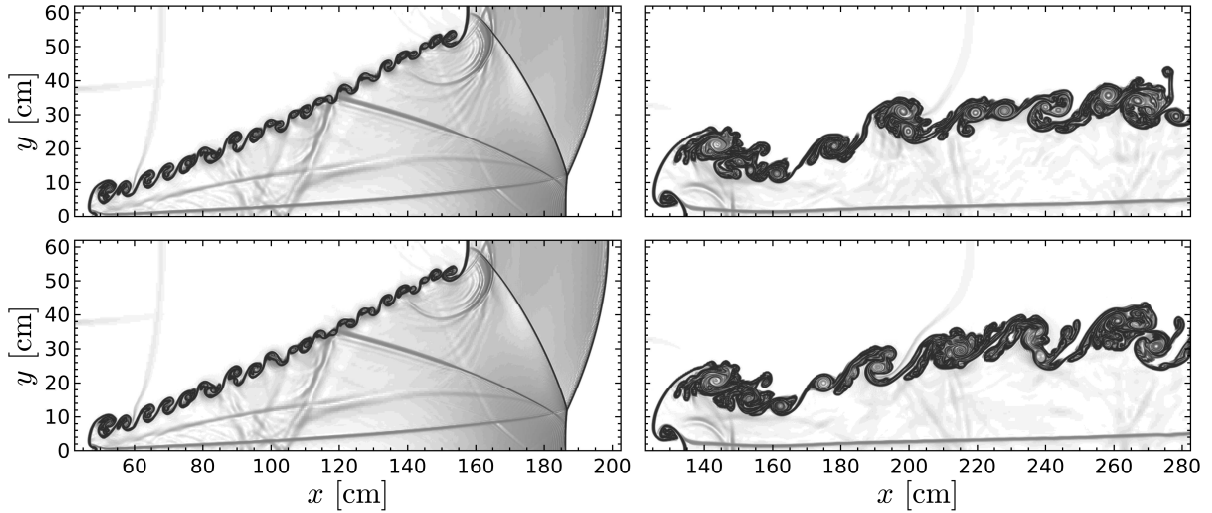}
                \caption{The morphology of the flow in the Hawley-Zabusky
                problem is shown with numerical schlieren density images for the
                baseline AMR solution (top row) and HAMR solution with $\kappa =
                1 \times 10^{-2}$ (bottom row). The panels in the left column
                show the morphology shortly after the shock passed through the
                interface ($t \approx 180\ \si{\second}$), while the structure
                of the interface at the final simulation time is shown in the
                right column. Note that the structure of the HAMR solution
                closely matches that of the baseline solution at early times.
                There are however discernible differences in the small scale
                structure of the mixed region at the final time. See text for
                details.}
                \label{fig:hz_combined}
            \end{figure}

            In our setup, the mesh is initially comprised of six blocks
            (consisting of $16 \times 16$ cells each) in the streamwise
            direction, and we allow $L=6$ levels of refinement. The MR tolerance
            is set to $\varepsilon = 1 \times 10^{-2}$, and we apply the solver
            adaptive scheme to the fluxes only, with a safety factor of $\kappa
            = 1 \times 10^{-2}$. The results corresponding to this choice of
            parameters are shown in \cref{fig:hz_combined}.
            The numerical schlieren density plot obtained with the baseline AMR
            scheme are shown in the top row, and those obtained with the HAMR
            scheme are shown in the bottom row at early (left column) and late
            times (right column) in the figure. As the shock passes through the
            interface and refracts, vorticity is deposited  along the interface
            (left column panels in \cref{fig:hz_combined}) resulting in
            extensive mixing at later times (right column panels in the figure).
            The conspicuous differences in the morphology of the flow at those
            late times is a consequence of strong sensitivity to perturbations,
            in this case introduced by the interpolation of fluxes in the HAMR
            solution.

        \subsection{Two-dimensional cellular detonation}
        \label{sec:cellular}

            The cellular detonation problem \cite{timmes2000b} is concerned with
            the growth and evolution of multidimensional instability of the
            detonation front.  We consider this problem as an essential example
            of the integrated multi-physics applications, with hydrodynamics now
            combined with a realistic EoS and a set of strongly coupled source
            terms. A relatively simple one-dimensional profile of the detonation
            wave is substantially altered if non-radial perturbations are
            introduced in multidimensional realizations of the problem. In this
            case, the energy delivered behind various parts of the initially
            planar detonation front will differ, resulting in nonuniform
            propagation speed of neighboring segments of the front.  Depending
            on the specific problem conditions, the perturbations organize into
            a set of separate regions that take the form of cell-like
            structures.

            Our setup for this problem matches that of \cite{timmes2000b}, with
            the exception of the nuclear network (we use a $7$ isotope network)
            and a twice narrower computational domain. The domain is covered
            with 20 blocks in the $x$-direction and one block in the
            $y$-direction, with the blocks each consisting of $16 \times 16$
            cells. The maximum number of mesh levels allowed is $L=7$, resulting
            in an effective model resolution of $1.25 \times 10^{-2}\ \si{\cm}$.
            This mesh refinement configuration is equivalent to approximately
            $160$ cells per burning length scale, ensuring that the problem
            dynamics are well resolved.  The refinement variables are density,
            pressure, and temperature, and the chosen MR tolerance of
            $\varepsilon = 6 \times 10^{-1}$ is sufficient to resolve the
            detonation front to the finest level of mesh resolution.

\begin{figure}
                \centering
                \includegraphics[width=0.86\textwidth]{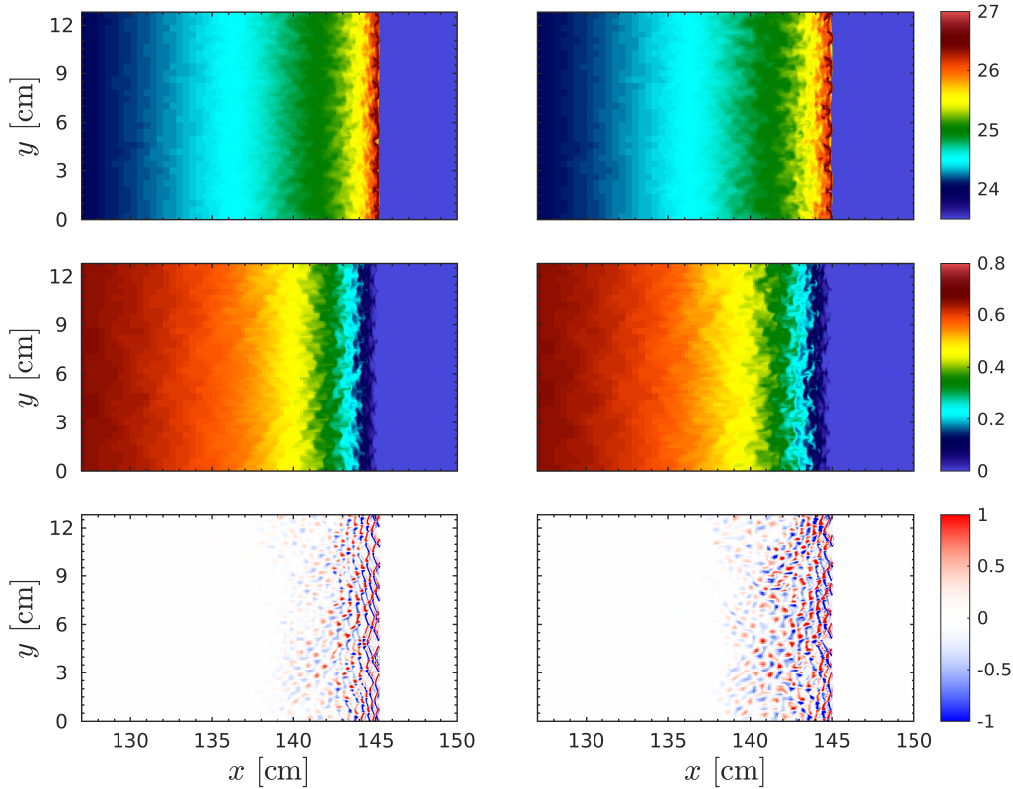}
                \caption{The morphology of the flow in quasi-steady state in the
                cellular detonation problem for the baseline AMR solution (left
                column panels) and the HAMR solution with $\kappa = 1 \times
                10^{-1}$ (right column panels). The individual rows of panels
                show nuclear energy generation rate (top row panels, log scale),
                silicon abundance (middle row panels), and the $z$-component of
                vorticity (bottom row panels). The vorticity is shown in units
                of $10^{9}\ \mathrm{s}^{-1}$. See text for details.}
                \label{fig:cellular_solutions}
            \end{figure}

            In \cref{fig:cellular_solutions}
            we show select components of the baseline AMR solution (left column
            panels) and the corresponding HAMR solution (right column panels)
            obtained with $\kappa = 1 \times 10^{-1}$ when the system is in
            quasi-steady state. In both solutions, the triple points can be seen
            as the local maxima in energy generation rate located along the
            leading shock front (top row panels in
            \cref{fig:cellular_solutions}). This energy release drives
            transverse waves which interact with the incident shock, resulting
            in a complex and inhomogeneous distribution of silicon (second row
            panels) in the post-detonation region. This process of compositional
            homogenization can be understood by examining the vorticity field
            (bottom row panels), which indicates the presence of strong
            rotational motions, and thus mixing. We note that the two solutions
            are qualitatively similar, however more diffusive structures are
            observed in the HAMR solution for the given safety factor. We defer
            a more detailed comparison of baseline and HAMR solutions for a
            range of safety factors to \cref{sec:hamraccuracy}.

       \subsection{Reactive turbulence}
       \label{sec:tburn}

            For the second integrated multi-physics application of the proposed
            method we select a reactive turbulence problem. Although this
            application shares several similarities with the cellular detonation
            problem introduced earlier, here the complex flow and action of
            source terms are no longer restricted to a narrow region of the
            problem domain. Furthermore, unlike in the case of detonation
            studies, turbulence is genuinely multi-dimensional in nature.  Also,
            even though the current study is limited to two spatial dimensions,
            all of the algorithmic components are examined in the same way as
            they would be in the three-dimensional case.

            The adopted reactive turbulence simulation setup closely matches
            that of \cite{brooker2020}, with the mesh being uniformly resolved
            with $1024$ cells per dimension.  Our model differs from that of
            Brooker et~al.\ in that turbulence is spectrally driven with the
            energy injected at a rate of $7 \times 10^{14}\ \text{ergs}\
            \text{s}^{-1}$, and the model is evolved toward quasi-steady state until
            $t \approx 125 \text{ms}$. The models used in the analysis are
            evolved for about one turnover time on the integral scale ($\approx
            40\ \text{ms}$) after the quasi-steady state is reached.

\begin{figure}
                \centering
                \includegraphics[width=0.81\textwidth]{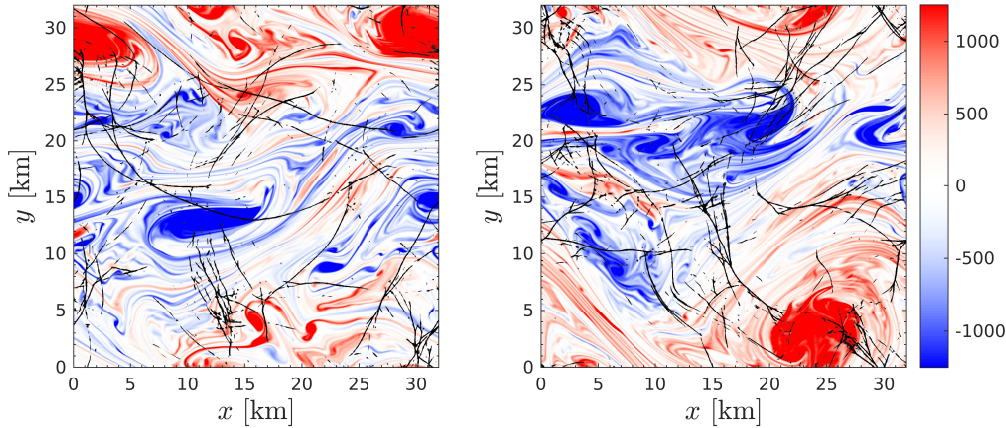}
                \caption{The vorticity field and divergence of velocity in the
                    two-dimensional turbulence model are shown with flooded
                    contours at $t = 125\ \text{ms}$\ (the beginning of
                    quasi-steady state) and $t = 165\ \text{ms}$\ (about one
                    turnover time on the integral scale into the quasi-steady state) in
                    the left and right panel, respectively. The solution in the
                    left panel was obtained using the reference scheme, and
                    provided initial conditions for the adaptive solution
                    obtained with $\tilde{\varepsilon} = 10^{-3}$, which is
                    shown in the right panel at the later time. The adaptive
                    solution displays qualitatively similar structures as the
                    reference solution, with two counter-rotating vortices
                    dominating the flow on the scale of the computational
                    domain. The narrow, extended structures correspond to the
                    velocity divergence smaller than $-800\ \text{s}^{-1}$,
                    associated with shocklets. Note that this particular
                    realization of the model does not include the effects of
                    nuclear burning.}
                \label{fig:tburn_vort}
            \end{figure}

            \Cref{fig:tburn_vort}
            shows the vorticity field and the velocity divergence in the
            quasi-steady state at the initial time for the reference model (left
            panel), and for the adaptive model at the final simulation time
            (right panel). We choose not to show the solutions at the same time
            because they do not significantly differ at the final time. Overall
            the adaptive solution displays similar flow structures to that of
            the reference solution, with the large scale flow dominated by
            counter-rotating vortices and numerous shocklets present. We note
            that in the quasi-steady state the average solution properties are
            expected to remain the same, which makes qualitative comparison of
            those two select models meaningful.
%
%
%
    %\section{Performance of the HAMR scheme}
    \section{Discussion}
    \label{sec:discussion}
%
%
%
        % preamble
        The introduction of the solver adaptive approach affects the solution
        accuracy and code performance to a varying degree depending on the
        problem at hand. These two factors are not independent of one another.
        For example, more frequent replacement of direct calculations with
        multiresolution-based interpolation, which offers computational savings,
        is enabled as the multiresolution threshold increases and solution
        accuracy is accordingly lowered (in general). In this section we discuss
        various aspects of this relation as well as other factors affecting
        computational performance of the new scheme.

        \subsection{Assessment of HAMR solution accuracy}
        \label{sec:hamraccuracy}

\begin{figure}
              \center
              \includegraphics[width=0.78\textwidth]{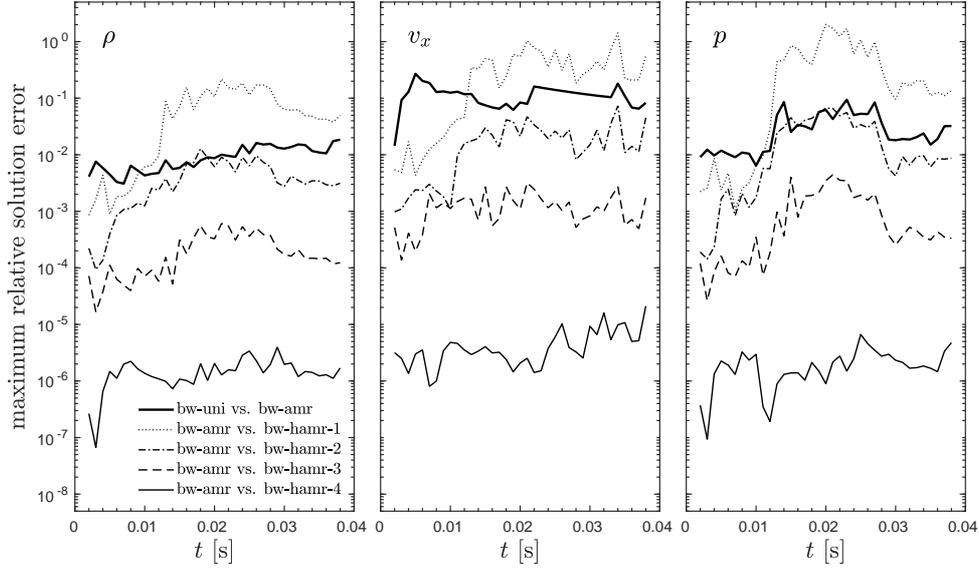}
              \caption{Maximum relative solution error for the interacting blast
                waves problem. The error at each time is calculated by
                normalizing the $L_{\infty}$ error norm by the maximum value of
                the corresponding variable obtained in the uniform mesh
                solution. The thick solid line shows the error evolution of the AMR
                solution with respect to the uniform mesh solution. The
                evolution of the errors obtained with the HAMR scheme with
                respect to the AMR solution are plotted with thin lines of differing
                styles as identified in the figure legend. The results are shown
                for density, velocity, and pressure for select values of the
                safety factor $\kappa$.  The safety factor is changing
                by a factor of $10$ between the HAMR models.}
              \label{fig:blast2_error}
            \end{figure}

            \paragraph{Interaction of two blast waves} To evaluate the impact
            of the solver adaptive procedure on the numerical solution, we first
            discuss the results obtained for the one-dimensional interacting
            blast waves problem (cf.\ \cref{sec:blast2}). We perform a set of
            simulations with the MR solver adaptive safety factor $\kappa$ being
            varied systematically from $10^{-4}$ to $10^{-1}$.
            \Cref{fig:blast2_error}
            shows the evolution of the normalized $L_{\infty}$ norm of the
            differences between the reference and adaptive numerical solutions.
            It is clear that the accuracy of the HAMR solutions (shown with the
            family of thin lines of differing styles in \cref{fig:blast2_error}),
            as compared with the AMR solution, scale approximately with the
            safety factor.  For example, the error in density (left panel in
            \cref{fig:blast2_error}) in the model \model{bw-hamr-1}, which
            corresponds to $\kappa = 1 \times 10^{-1}$, reaches a value of about
            $0.2$, while for the model \model{bw-hamr-4}, which corresponds to
            $\kappa = 1 \times 10^{-4}$, the error never exceeds a value of $1
            \times 10^{-5}$.

            Although the errors in velocity and pressure (middle and right
            panels in \cref{fig:blast2_error}, respectively) slightly exceed the
            density error, the overall scaling with the safety factor appears to
            hold. This behavior is consistent with the MR analysis provided by
            \cite{harten1994}.  We find these results encouraging because
            despite the intrusive character of the solver-adaptive approach and
            highly nonlinear character of the problem, the solution trajectories
            do not show any inconsistent departures from the reference solution.

            One can also quantify the error in the AMR solution by using the
            uniform mesh as the reference solution. This error, shown with the
            thick solid line in \cref{fig:blast2_error}, remains within the
            expected ranges (recall that the tolerance for mesh adaptation is
            $\varepsilon = 1 \times 10^{-2}$) for each of the variables
            throughout the simulation. The error in the HAMR models with $\kappa
            = 1 \times 10^{-4}$ through $\kappa = 1 \times 10^{-2}$ remain
            smaller than this, implying that the solver adaptive procedure is
            not increasing the overall solution error for these values of the
            safety factor.

            \paragraph{Hawley-Zabusky problem} From the perspective of error
            control, the regions of interest in the HZ problem may appear to be
            limited to the salient solution features, in particular the shocks
            and vortices. However, the background flow is in fact extremely rich
            in this case (we refer the interested reader to \cite{zabusky1999}
            for a more detailed description of those flow structures).

\begin{figure}
                \centering
                \includegraphics[width=0.70\textwidth]{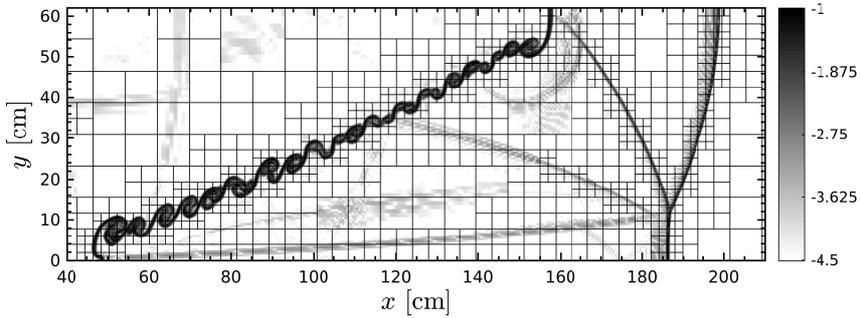}
                \caption{The log of MR detail coefficients on the finest
                  level of each the local MR hierarchy on each AMR block is
                  plotted at time $t \approx 180 \si{\second}$.  At the selected
                  MR tolerance of $\varepsilon = 1 \times 10^{-2}$, the
                  algorithm allocates blocks at the finest AMR level ($L=6$) in
                  two main regions: along the material interface where vorticity
                  is large, and along the passing shock waves, where pressure
                  gradients are large. Near the important solution features
                  there are regions where the mesh is relatively fine (due to
                  the block-based AMR format) but the MR detail coefficients are
                  relatively small, inviting the use of the solver adaptive
                  scheme.}
                \label{fig:hz_mesh}
            \end{figure}

            \Cref{fig:hz_mesh}
            shows the field of MR detail coefficients at $t \approx 180
            \si{\second}$, with the AMR block outlines overlaid. When compared
            with the numerical schlieren image at the same time (see
            \cref{fig:hz_combined}), it is evident that the MR detail
            coefficients closely track the prominent solution features. The
            coefficients associated with the leading shock, reflected shock, and
            Mach stem reach their maximal values in excess of $0.1$. These high
            values force refinement to the finest available AMR level.
            Meanwhile the weaker acoustic waves reverberating behind the shocks
            and across the interface result in significantly smaller values of
            detail coefficients, necessitating less mesh refinement. These
            values are below the specified MR tolerance, and with the adopted
            safety factor in this particular model of $\kappa = 1 \times
            10^{-2}$, the corresponding parts of the flow are subject to the
            solver adaptive scheme.

            One of the key quantities characterizing the HZ problem is the
            vorticity, which is the result of the baroclinic source term driven
            by the interaction of density and pressure gradients. One can expect
            that the amount of vorticity produced will sensitively depend on the
            accuracy of the numerical scheme, and in particular on its ability
            to describe solution gradients. Likewise the amount of vorticity
            produced will depend on the prescribed MR tolerance. Also the
            integrated vorticity better characterizes the overall performance of
            the solver than, for example, the structure of the interface. This
            is because the growth of instability across the interface in this
            problem is limited only by numerical diffusion and thus does not
            converge under mesh refinement.

\begin{figure}
                \centering
                \includegraphics[width=0.72\textwidth]{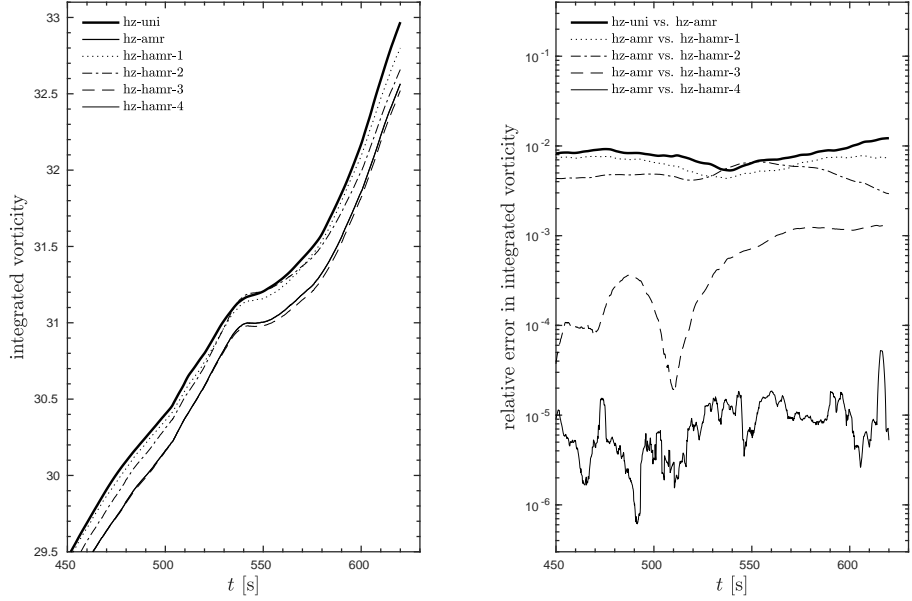}
                \caption{The time evolution of integrated vorticity and its relative
                error in the Hawley-Zabusky problem at late times. The
                integrated vorticity evolution is shown in the left panel for
                the reference uniform mesh solution (thick solid line), AMR
                solution (medium solid line), and a set of HAMR models (family
                of thin lines of differing styles). The time evolution of the
                relative error in integrated vorticity is shown in the right
                panel during the same time window for each HAMR solution. See
                text for details.}
                \label{fig:hz_vort_error}
            \end{figure}

            \Cref{fig:hz_vort_error}
            shows the evolution of integrated vorticity and its relative error
            for our set of HZ models, where the safety factor varies between $1
            \times 10^{-4}$ and $1 \times 10^{-1}$. The vorticity during the
            late time discussed here gradually increases, with a temporary
            plateau around $t \approx 540\ \si{\second}$ due to a vortex pair
            merger. The amount of vorticity produced in the AMR model,
            \model{hz-amr}, which serves as the baseline model for the HAMR
            solutions, is smaller in comparison to the uniform mesh reference
            model, \model{hz-uni}, by nearly $1\%$, but closely follows the
            general trend. The HAMR model with the least restrictive safety
            factor of $\kappa = 1 \times 10^{-1}$, \model{hz-hamr-1}, shows the
            largest deviation away from the baseline \model{hz-amr} model, but
            again the overall trend in the vorticity evolution is preserved. As
            the solver adaptive tolerance is tightened, the vorticity in the
            consecutive \model{hz-hamr} models converge to the baseline
            \model{hz-amr} model.

            \paragraph{Two-dimensional cellular detonation} In the cellular
            detonation problem the goal is to preserve the cellular structure of
            the detonation and accurately capture the nuclear energy generation
            rate, composition of the combustion products, and also the
            production of vorticity, which is responsible for mixing in the
            reaction zone behind the front. The last element is also of
            importance for incomplete burning taking place in the tail of the
            reaction zone. We note that depending on the adopted threshold this
            extended reaction zone is only partially resolved.

\begin{figure}
                \centering
                \includegraphics[width=0.7\textwidth]{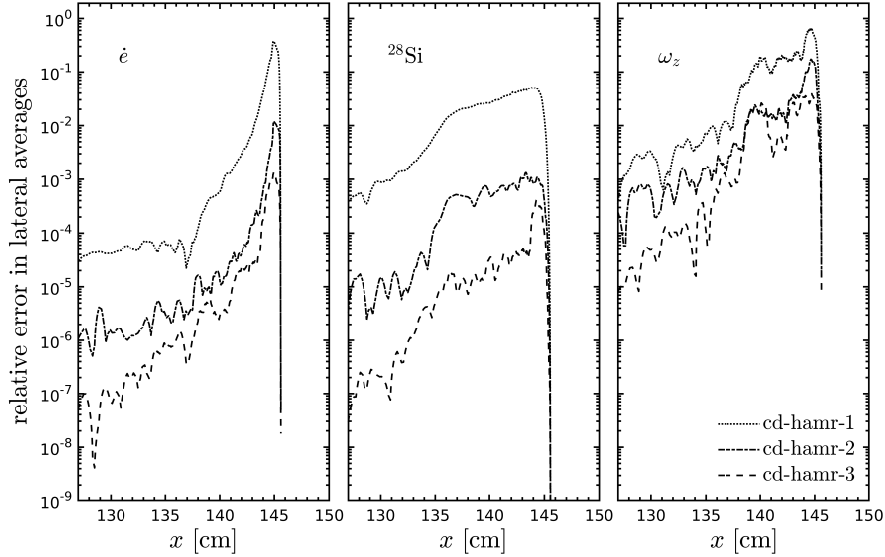}
                \caption{Errors in lateral averages of energy generation rate,
                    abundance of silicon, and vorticity in the reaction zone for the
                    cellular detonation problem. The solver adaptive approach is
                    applied to the hydrodynamic fluxes, Helmholtz EoS, and the
                    reactive source term, and the safety factor ranges from $\kappa
                    = 10^{-3}$ to $\kappa = 10^{-1}$.}
                \label{fig:lateralavg_error}
            \end{figure}

            In our study, the safety factors range from $1 \times 10^{-3}$ to $1
            \times 10^{-1}$. \Cref{fig:lateralavg_error}
            shows the $L_{\infty}$ error norm in lateral averages of energy
            generation rate, silicon abundance, and vorticity at the final
            simulated time. The respective errors are normalized by the
            $L_{\infty}$ value obtained in the reference AMR model,
            \model{cd-ref}. It is evident from the left panel of \cref{fig:lateralavg_error}
            that in the case of the \model{cd-hamr-1} model, the maximal error
            in energy generation rate is approximately equal to the adopted AMR
            tolerance of $\varepsilon = 1 \times 10^{-1}$. This implies that the
            error in this quantity due to the solver adaptive scheme is not
            dominating the overall solution error.

            The middle panel of \cref{fig:lateralavg_error} shows the
            distribution of silicon, which is the main product of burning in
            this problem. The error in this quantity is well controlled in all
            HAMR models. The error in vorticity, shown in the right panel of
            \cref{fig:lateralavg_error}, is the greatest in the model
            \model{cd-hamr-1}, and exceeds that of the AMR tolerance ($\approx
            0.8$ vs.\ $0.6$).  However, lowering the safety factor produces much
            more satisfactory results in both energy generation rate and
            vorticity. One item to note is the crossing of error curves in
            vorticity between the models \model{cd-hamr-2} and \model{cd-hamr-3}
            in the trailing edge of the reaction zone ($\approx 140$\ cm). We
            speculate that one possible reason for that behavior is the highly
            unsteady character of the solution, which naturally results in
            unsteady behavior of errors.

\begin{figure}
                \centering
                \includegraphics[width=0.45\textwidth]{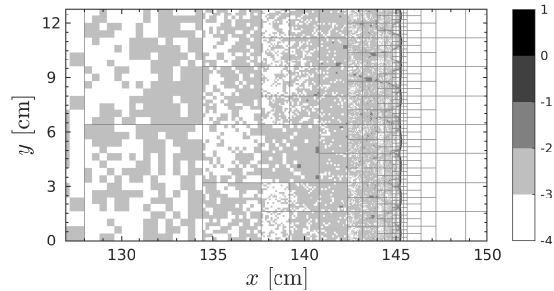}
                \caption{Mesh structure and MR detail coefficients for the cellular
                    detonation problem in the \model{cd-ref} model. The mesh structure
                    is indicated by the outlines of the individual mesh blocks
                    (each block contains $16 \times 16$ cells). The detail
                    coefficients are shown in the log scale with flooded
                    contours. Note that upstream of the detonation front the
                    values of the detail coefficients are zero.}
                \label{fig:cellular_mesh}
            \end{figure}

            The primary source of solution errors in this problem is the leading
            shock front, the associated reaction zone, and the detonation cells.
            These features are clearly captured by the MR decomposition, as
            illustrated by the distribution of detail coefficients in
            \cref{fig:cellular_mesh}.
            Also, it should be noted that the interpolation of fluxes, EoS
            outputs, and source terms is relatively active inside the reaction
            zone in the \model{cd-hamr-1} model compared to models with a
            smaller value of the safety factor. This is not unexpected given
            that this region contains pockets of relative smoothness between the
            detonation cells and also in the detonation cell interiors. An
            example is the intercell region near $(x,y) \approx (145, 9)$ and
            the detonation cell interior near $(x,y) \approx (145, 11)$. In
            effect, in the coarse model \model{cd-hamr-1}, the highly nonlinear
            part of the solution dominated by burning is fed with a significant
            perturbation that is due to MR interpolation.  However as soon as
            the energy generation rate is added to the set of mask indicator
            variables, the solver adaptive mask covers the entire reaction zone,
            and consequently no interpolation is performed. Overall the
            distribution of detail coefficients correctly reflects on the
            smoothness of the flow downstream of the detonation front, resulting
            in a gradual decrease in mesh resolution.

\begin{figure}
                \centering
                \includegraphics[width=0.64\textwidth]{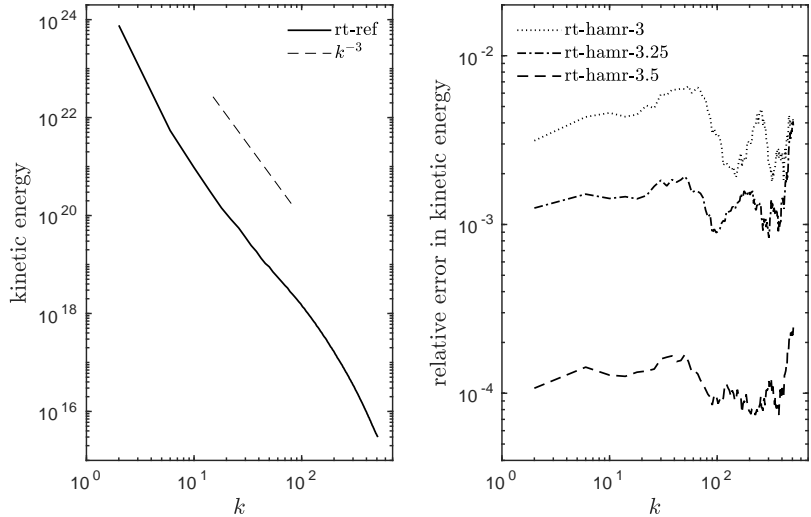}
                \caption{The turbulent kinetic energy spectra in the reactive
                    turbulence problem. The energy spectrum in the reference
                    non-adaptive solution as a function of wavenumber,
                    $k$, is shown in the left panel along with the Kraichnan
                    $k^{-3}$ scaling law, indicated with a straight line
                    segment. The relative error for each of the solver adaptive
                    models is shown in the right panel. The solver adaptive
                    tolerances range from $\tilde{\varepsilon} = 10^{-3.5}$ to
                    $\tilde{\varepsilon} = 10^{-3.0}$. Note that the error
                    values are consistent with the chosen tolerance of the
                    adaptive scheme.}
                \label{fig:tburn_kespec}
            \end{figure}

            \paragraph{Reactive turbulence} One of the fundamental statistics
            used to characterize isotropic, homogeneous turbulence is the
            kinetic energy spectrum. The theory of turbulence predicts a
            universal scaling law in the intertial range, with kinetic energy of
            velocity fluctuations (KE) to scale as $k^{-3}$ in two-dimensional
            situations, as predicted by Kraichnan \cite{kraichnan1967}. It is
            reproducing this behavior in numerical experiments that provides
            elemental confidence in simulation outcomes. \Cref{fig:tburn_kespec}
            shows the kinetic energy spectra (left panel) and respective errors
            (right panel) in our set of numerical experiments. The errors in the
            solver adaptive solutions were calculated with respect to the
            reference, non-adaptive model. The overall shape of the kinetic
            energy spectra matches closely that of the theoretical $k^{-3}$
            scaling relation. As in the previous test cases, the errors in the
            KE spectra obtained with the solver adaptive solutions appear to
            vary in a way consistent with the changes in the solver adaptive
            tolerance, $\tilde{\varepsilon}$. However, we were unable to obtain
            satisfactory solutions with tolerances greater than
            $\tilde{\varepsilon} = 1 \times 10^{-3}$. In such heavily
            interpolated models the solution structure displayed numerical
            artifacts in the form of undershoots and overshoots in the solution
            components. Furthermore, a relatively small tightening of the
            tolerance resulted in a disproportionately large improvement in the
            solution quality.  We speculate that this behavior might be
            attributed to thermodynamically inconsistent MR interpolation of the
            EoS. The thermodynamic consistency of the Helmholtz EoS was shown by
            Timmes \& Swesty \cite{timmes2000a} to be of crucial importance for
            obtaining physically relevant solutions.

            In the previously discussed problems our focus was not necessarily
            focused on the primitive solution components but rather on the
            solution functionals, as they are of main interest from the
            application point of view. Some examples here are the vorticity in
            the Hawley-Zabusky problem, and the production of silicon in the
            cellular detonation problem. Likewise here our focus is on the
            ignition time, $\tau_{\mathrm{ig}}$. This quantity describes the
            potential of the physical system to develop deflagrations and
            detonations, which qualitatively change the system evolution.
            Therefore it is particularly important for the numerical scheme to
            correctly describe a part of the solution containing the shortest
            ignition times.

\begin{figure}
                \centering
                \includegraphics[width=0.70\textwidth]{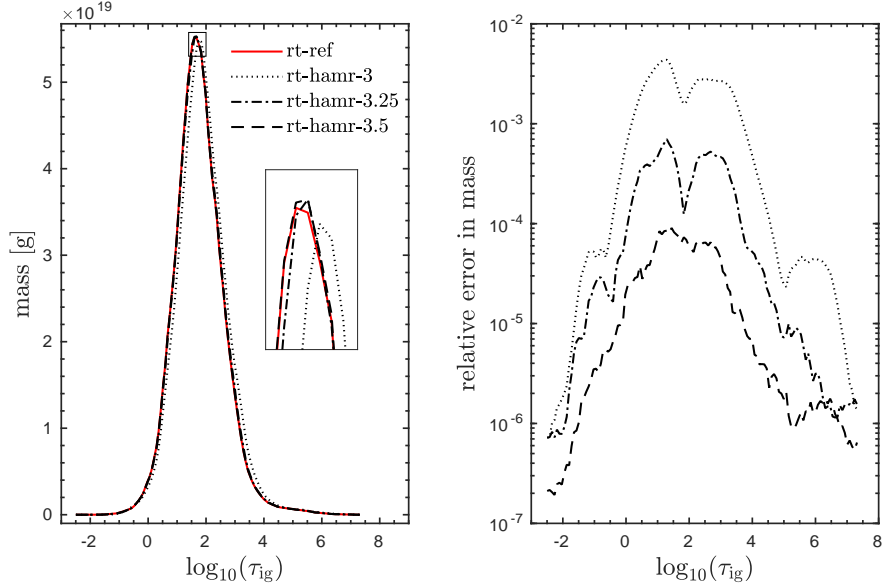}
                \caption{The mass distribution as a function of ignition times
                    in the reactive turbulence problem is shown for a set of
                    adaptive solutions (left panel). The solver adaptive
                    tolerances range from $\tilde{\varepsilon} = 10^{-3.5}$ to
                    $\tilde{\varepsilon} = 10^{-3.0}$, with $\tilde{\varepsilon}
                    = 0$ corresponding to the reference case. The inset shows
                    in detail the mass distributions near their peak values. The
                    relative error in mass distribution is shown in the right
                    panel. The line styles used to plot the relative error
                    correspond to the legend in the left panel. Note that the
                    mass distributions as well as their respective errors show
                    systematic deviations with relatively less mass at short
                    ignition times as the error tolerance becomes less
                    stringent.}
                \label{fig:tburn_igtm}
            \end{figure}

            One way of characterizing various populations of ignition times is
            to construct a histogram of mass as a function of ignition time.
            Such a histogram is shown for our set of reactive turbulence models
            in the left panel of \cref{fig:tburn_igtm}.
            Although it is encouraging to see the overall shape of the
            distributions obtained from adaptive models closely matching the
            reference distribution, a closer inspection of histogram data
            reveals some systematic differences. The inset plot in the left
            panel in the figure shows a closeup view of the region near the
            peaks of the distributions. We observe the increasing amount of mass
            at longer ignition times as the tolerance $\tilde{\varepsilon}$ is
            progressively more relaxed. This is confirmed by the examination of
            differences of mass distributions obtained from solver adaptive
            models with that of the reference model (right panel in
            \cref{fig:tburn_igtm}).  The observed behavior can be understood as
            the result of greater numerical dissipation in more heavily
            interpolated models, which results in the decrease of amplitudes of
            turbulent fluctuations and therefore the decrease of the number of
            fluid parcels with short ignition times.  From the application point
            of view, the consequence of the added dissipation is that the
            deflagration ignition process or the process of transition to
            detonation may not be truthfully described in heavily interpolated
            models. This is consistent with our earlier conclusion that in this
            application in which there is an extremely strong dependence of the
            source term on temperature, the numerical convergence is problematic
            and demands good control over the error.

        \subsection{Computational efficiency}

            To analyze the savings offered by the solver adaptive scheme in
            terms of computational cost, Harten \cite{harten1994} introduced the
            efficiency factor, which compares the number of cells used to obtain
            the non-adaptive model to the number of cells used to obtain the
            corresponding MR model. We adapt Harten's definition to be
            compatible with block-structured AMR format as follows. The number
            of cells in the mesh at timestep $n$ is $|\bm{\mathcal{B}}^{n}|
            N_{\mathcal{L}}$, where $|\bm{\mathcal{B}}^{n}|$ is the total number
            of leaf blocks, and the total number of cells in the adaptive mask
            over the collection of blocks is
            $\sum_{b=1}^{|\bm{\mathcal{B}}^{n}|} N_{1} + |\bm{\mathcal{M}}_{b}^{n}|$,
            where $|\bm{\mathcal{M}}_{b}^{n}|$ is the number of cells included in the
            local mask of block $b$. With these terms introduced, we define the
            efficiency factor to be the ratio of the total number of cells used
            in the simulation to the total number of adaptive evaluations,
            \begin{equation}
                \eta = \frac{\sum_{n = 1}^{n_{\Delta t}} |\bm{\mathcal{B}}^{n}|
                    N_{\mathcal{L}}}{\sum_{n = 1}^{n_{\Delta t}}
                    \left( \sum_{b=1}^{|\bm{\mathcal{B}}^{n}|} N_{1} +
                    |\bm{\mathcal{M}}_{b}^{n}| \right)},
            \end{equation}
            where $n_{\Delta t}$ is the total number of timesteps. Note that the
            so-defined efficiency factor provides the maximal estimate of
            possible savings and typically provides an upper bound for the
            reduction in time-to-solution of computer simulations.

            To enable a more direct comparison between solver components, we
            define the efficiency factors with respect to each module as
            $\eta_{\mathrm{hydro}}$, $\eta_{\mathrm{eos}}$, and
            $\eta_{\mathrm{burn}}$.  These will differ from one another due to
            the unique attributes of each module. For example, the reactive
            source term is not activated unless the density and temperature are
            sufficiently high.

            To measure improvements in execution times, we adopt the speedup
            factor of Bihari \cite{bihari1999},
            \begin{equation}
                \zeta = \frac{\mathrm{runtime\ per\ call\ without\
                    solver\ adaptivity}}{\mathrm{runtime\ per\ call\ with\
                    solver\ adaptivity}}.
            \end{equation}
            The runtime is measured on a per-call basis. This is because the
            trajectory of the solution obtained with the solver adaptive scheme
            may differ from the trajectory of the solution obtained without
            solver adaptivity. In consequence the AMR refinement pattern may
            differ as well, requiring a different number of calls to the
            individual solver components.

            As in the case of the efficiency factor, we calculate speedup
            factors for the individual solver components, labeling the factors
            as $\zeta_{\mathrm{hydro}}$, $\zeta_{\mathrm{eos}}$, and
            $\zeta_{\mathrm{burn}}$. We reiterate that these speedup factors
            will in general be smaller than the corresponding efficiency factors
            defined earlier.

            \paragraph{Interaction of two blast waves} For the problem of two
            interacting blast waves, the HAMR scheme achieves a significant
            reduction in the number of fluxes evaluated, while maintaining
            solution accuracy compared with the baseline \model{bw-amr} model
            (see \cref{sec:hamraccuracy}). \Cref{tbl:bw}
            \begin{table}
                \centering
                {\footnotesize
                \caption{Computational efficiency data for the interacting blast
                    waves problem. The efficiency factor, $\eta_{\mathrm{hydro}}$, and
                    the reduction in the number of fluxes compared with the
                    baseline model \model{bw-amr} are shown as a function of the
                    safety factor, $\kappa$.
                    \label{tbl:bw}}
                    %\begin{tabular}{SSSSSSSSS} \toprule
                    \begin{tabular}{lll} \hline
                        {$\kappa$} & {$\eta_{\mathrm{hydro}}$} &
                            {fraction of fluxes} \\ \hline
                        {$10^{-1}$} & 1.46 & 0.68 \\
                        {$10^{-2}$} & 1.29 & 0.78 \\
                        {$10^{-3}$} & 1.20 & 0.83 \\
                        {$10^{-4}$} & 1.18 & 0.85 \\
                        \hline
                    \end{tabular}
                }
            \end{table}
            provides computational efficiency data, which includes the
            efficiency factor and information about the reduction in the
            number of fluxes computed compared to the model bw-amr. The
            collected data indicates that for models that we deem acceptable
            from the solution accuracy point of view, one may expect
            efficiency factors between about $1.2$ to $1.3$.  Relative to
            the baseline AMR model, \model{bw-amr}, the fraction of
            evaluated fluxes for these models is about $0.8$. We conclude
            that for this problem dominated by a relatively few strong,
            localized discontinuities always resolved to the finest AMR
            level, the scheme achieves a performance improvement of roughly
            $20 \%$.

            \paragraph{Hawley-Zabusky problem} In the HZ problem the mesh filling
            factor is significantly greater than in the interacting blast
            waves problem due to the mixing of fluid along the interface. For
            this reason one may expect less performance gain from the HAMR
            approach. The data in \cref{tbl:hz}
            % hawley zabusky speedup table for L = 6 and L = 7 amr levels
            \begin{table}
                \centering
                {\footnotesize
                \caption{Efficiency factor, $\eta_{\mathrm{hydro}}$, and
                    speedup factors, $\zeta_{\mathrm{intrfc}}$,
                    $\zeta_{\mathrm{states}}$, $\zeta_{\mathrm{rieman}}$, for the
                    Hawley-Zabusky problem with $L = 6$ and $L=7$ levels of
                    refinement, and an AMR tolerance of $\varepsilon = 0.01$,
                    respectively. A range of safety factors are tested, with the
                    baseline AMR model corresponding to $\kappa = 0$.
                    \label{tbl:hz}}
                    %\begin{tabular}{SSSSSSSSS} \toprule
                    \begin{minipage}{8cm}
                    \begin{tabular}{llllllll} \hline
                        {$L$} & {$\kappa$} & {$\eta_{\mathrm{hydro}}$} &
                            {$\zeta_{\mathrm{intrfc}}$} & {$\zeta_{\mathrm{states}}$} &
                            {$\zeta_{\mathrm{rieman}}$} & {$\zeta_{\mathrm{hydro}}$} &
                            {$t_{\mathrm{wall}} (\si{\hour})$}\footnote{Elapsed real time, measured in hours.} \\ \hline
                        {$6$} & {$10^{-1}$} & 1.37 & 1.14 & 1.22 &
                            1.26 & 1.20 & 3.58 \\
                        {$6$} & {$10^{-2}$} & 1.15 & 1.08 & 1.10 &
                            1.12 & 1.10 & 3.71 \\
                        {$6$} & {$10^{-3}$} & 1.06 & 1.02 & 1.03 &
                            1.04 & 1.03 & 3.74 \\
                        {$6$} & {$10^{-4}$} & 1.04 & 1.02 & 1.03 &
                            1.03 & 1.03 & 3.74 \\
                        {$6$} & {$0$} & \textendash & \textendash &
                            \textendash & \textendash & \textendash & 3.76 \\
                        \hline
                        {$7$} & {$10^{-1}$} & 1.34 & 1.14 & 1.19 &
                            1.22 & 1.18 & 19.41 \\
                        {$7$} & {$10^{-2}$} & 1.16 & 1.07 & 1.08 &
                            1.09 & 1.08 & 21.02 \\
                        {$7$} & {$10^{-3}$} & 1.06 & 1.01 & 1.02 &
                            1.03 & 1.02 & 21.11 \\
                        {$7$} & {$10^{-4}$} & 1.03 & 1.01 & 1.02 &
                            1.02 & 1.02 & 21.36 \\
                        {$7$} & {$0$} & \textendash & \textendash &
                            \textendash & \textendash & \textendash & 21.46 \\
                        \hline
                    \end{tabular}
                    \end{minipage}
                }
            \end{table}
            confirms this expectation. In particular, the obtained efficiency
            factors are around $1.1$ for models with a still acceptable quality
            ($\kappa$ between $1 \times 10^{-3}$ and $1 \times 10^{-2}$). Also,
            the speedup factors for the individual components of the hydro
            solver are typically less than $1.1$ for that range of models.
            Furthermore the performance gains are less for \texttt{intrfc}
            than for \texttt{states} and \texttt{rieman} modules because of
            the poorer data locality for the (interface) reconstruction part
            of the PPM algorithm. Finally, we speculate that the improvement in
            simulation wallclock time is probably hindered by the
            exacerbated load imbalance due to a difference in computational
            cost between blocks where some fraction of fluxes were
            interpolated, and those blocks where no such interpolation was
            used. The set of results obtained with mesh resolution increased by a
            factor of $2$ ($L = 7$ in the table) displays similar trends,
            although the gains are somewhat smaller perhaps due to the increased
            complexity of the small scale structure.

            \paragraph{Cellular detonation problem} The efficiency data for the
            cellular detonation problem is shown in \cref{tbl:cd}.

            % cellular speedup table for L = 7 and L = 8 amr levels
            \begin{table}
                \centering
                {\footnotesize
                \caption{Computational efficiency data for the cellular
                    detonation problem. The table shows the efficiency factors,
                    $\eta_{\mathrm{hydro}}$, $\eta_{\mathrm{eos}}$, and
                    $\eta_{\mathrm{burn}}$, and speedup factors,
                    $\zeta_{hydro}$, $\zeta_{eos}$, and $\zeta_{burn}$ for solutions
                    obtained with with $L = 7$ and $L=8$ levels of refinement.
                    The chosen AMR tolerance for these models is $\varepsilon =
                    0.6$. Results for several runs obtained with a range of
                    safety factors are presented with the baseline AMR model
                    corresponding to $\kappa = 0$. \label{tbl:cd}}
                    %\begin{tabular}{SSSSSSSSS} \toprule
                    \begin{tabular}{lllllllll} \hline
                        {$L$} & {$\kappa$} & {$\eta_{\mathrm{hydro}}$} &
                            {$\eta_{\mathrm{eos}}$} & {$\eta_{\mathrm{burn}}$} &
                            {$\zeta_{\mathrm{hydro}}$} &
                            {$\zeta_{\mathrm{eos}}$} & {$\zeta_{\mathrm{burn}}$}
                            & {$t_{\mathrm{wall}} (\si{\hour})$} \\ \hline
                        %{$7$} & {$10^{-0.5}$} & 1.642 & 2.807 & 2.283 &
                        %  1.208 & 1.346 & 1.305 & 2.569 \\
                        {$7$} & {$10^{-1}$} & 1.47 & 2.14 & 1.55 &
                            1.15 & 1.25 & 1.18 & 2.57 \\
                        %{$7$} & {$10^{-1.50}$} & 1.35 & 1.72 & 1.17 &
                        %    1.12 & 1.16 & 1.06 & 2.56 \\ 
                        {$7$} & {$10^{-2}$} & 1.31 & 1.58 & 1.05 &
                            1.12 & 1.14 & 1.02 & 2.60 \\
                        %{$7$} & {$10^{-2.50}$} & 1.30 & 1.51 & 1.00 &
                        %    1.11 & 1.12 & 0.99 & 2.63 \\
                        {$7$} & {$10^{-3}$} & 1.30 & 1.51 & 1.00 &
                            1.11 & 1.11 & 0.99 & 2.62 \\
                        {$7$} & {$0$} & \textendash & \textendash &
                            \textendash & \textendash & \textendash & \textendash &
                            2.64 \\
                        \hline
                        %{$8$} & {$10^{-0.5}$} & 1.652 & 2.846 & 2.356 &
                        %    1.195 & 1.323 & 1.365 & 5.329 \\
                        {$8$} & {$10^{-1}$} & 1.55 & 2.12 & 1.57 &
                            1.15 & 1.23 & 1.19 & 5.55 \\
                        %{$8$} & {$10^{-1.50}$} & 1.33 & 1.70 & 1.20 &
                        %    1.12 & 1.15 & 1.08 & 5.77 \\
                        {$8$} & {$10^{-2}$} & 1.29 & 1.55 & 1.08 &
                            1.11 & 1.12 & 1.02 & 5.89 \\
                        %{$8$} & {$10^{-2.50}$} & 1.27 & 1.49 & 1.03 &
                        %    1.10 & 1.10 & 1.00 & 6.00  \\
                        {$8$} & {$10^{-3}$} & 1.27 & 1.46 & 1.01 &
                            1.11 & 1.10 & 0.99 &  5.95 \\
                        {$8$} & {$0$} & \textendash & \textendash &
                            \textendash & \textendash & \textendash & \textendash &
                            5.99 \\
                        \hline
                    \end{tabular}
                }
            \end{table}
            The data now include two additional physics describing the
            thermodynamics via a complex EoS, and the reactive source term
            for nuclear energy generation. The hydrodynamics module shows
            somewhat better performance compared to the two pure
            hydrodynamics problems due to the heavily interpolated ambient
            medium upstream of the detonation front. The same level of
            performance oberserved for the hydrodynamics is also observed in the
            EoS module. This kind of superficial performance improvement does
            not apply to burning however, because the code omits calculation of
            the nuclear burning source term if the temperature is too low, as is
            the case in the cold fuel region upstream of the shock. Furthermore,
            the MR analysis does not allow any interpolation inside the reaction
            zone for values of $\kappa$ below $1 \times 10^{-2}$. For this
            reason, we observe no speedup ($\zeta_{\mathrm{burn}} = 0.99$) for
            the model with $\kappa = 1 \times 10^{-3}$. Similar trends are
            observed in the higher resolution ($L = 8$) model.

            \paragraph{Reactive turbulence problem} Unlike in the previous
            problems discussed, the turbulence problem is characterized by the
            presence of substantial flow structure occupying the entire domain,
            and on all scales. This is reflected by the relatively narrow range
            of solver adaptive tolerances that we considered for this problem
            (cf.\ \cref{sec:tburn}), where we found that for coarse tolerances
            the solution appeared polluted by numerical artifacts, while there
            was no interpolation performed for tolerances of about an order of
            magnitude tighter. In consequence, the solver adaptive approach was
            enabled only in relatively small parts of the solution. We found
            that for the model \model{rt-hamr-3} ($\tilde{\varepsilon} = 1
            \times 10^{-3}$), the speedup factors were at most a few percent for
            the reactive source term, while there were no noticeable gains for
            the hydro or EoS solver components. For the remaining models, no
            speedup was observed for any of the solver components. We conclude
            that for turbulence models, the strong coupling between the
            thermodynamically sensitive equation of state and nonlinear
            hydrodynamics appears to make the problem effectively intractable
            from the point of view of solver-adaptivity.

        \subsection{Overall performance considerations}
        \label{sec:performance}

            We considered a set of applications ranging from a shock-dominated
            one-dimensional flow problem to complex multi-dimensional,
            multi-physics situations using the proposed HAMR scheme. In each of
            these applications, the domain is occupied by a small number of
            discontinuous flow features (e.g.\ shocks) against a rich background
            (e.g.\ reflected acoustic waves). Furthermore, the prominent flow
            features interact with each other frequently and create new,
            complicated structures. With respect to these situations the
            discussion of the overall performance of the HAMR scheme naturally
            turns to two main points: the efficiency of the MR-driven AMR mesh,
            and the deviation of the solver adaptive solution trajectory from
            the reference solution trajectory.

            % efficiency
            We find that the filling factor at the block level is what primarily
            determines the gains offered by adding solver adaptivity to the
            MR-driven AMR approach. This is in contrast to the classic AMR
            computations in which the computational efficiency is determined by
            the overall mesh filling factor (cf.\ \cref{sec:introduction}). In
            situations where the block filling factor is high, the solver
            adaptive scheme typically offers little reduction in computational
            complexity with essentially no savings in computational time.  This
            issue is evident in the problem of reactive turbulence, for which we
            did not observe a significant gain from solver adaptivity. However,
            one can anticipate that such gains would be offered in the case of
            inhomogeneous turbulence, where the solution contains, for example,
            a flame front (discontinuous structure) that separates fuel from
            burning products. In particular, studies of deflagrations in the
            distributed burning regime appear especially suitable for the HAMR
            approaches.

            % propagation of errors
            In general we expect modeling of turbulence with the solver adaptive
            approach to be challenging. This is because, as we observed in an
            extended set of reactive turbulence simulations, the accumulation of
            interpolation error caused progressive deviation of the
            solver adaptive solutions away from the reference solution. That is
            we observed significant differences between the solutions at late
            times. Therefore, the solver adaptive approach may not be suitable
            for problems expected to produce new phenomena after a transient
            period of evolution. We note that much the same accumulation of
            error is expected to occur in pure AMR simulations as well, for
            example in studies of homogeneous turbulence where the mesh blocks
            are continuously created and destroyed depending on the local
            smoothness of the solution.

            We note that our set of application examples are more typical of
            basic physics or discovery science studies rather than engineering
            problems. The former class of problems, for example, are frequently
            characterized by very high Reynolds numbers and thus prone to produce
            increasing amounts of structure in the solution as mesh resolution
            increases. A representative example from our application studies is
            the HZ problem, in which the small scale structure in the mixed
            region is limited only by the numerical dissipation of the solver.
            Had the HZ problem included sub-grid scale viscous dissipation
            models typically used in engineering applications, the amount of
            small scale structure would be limited, with the results converging
            upon adequate mesh refinement. In such an engineering model, one
            could expect that the HAMR scheme would be increasingly more
            efficient, as the solution would be progressively more interpolated
            due to an increase in the amount of solution regions with sufficient
            smoothness.

            Also, there might be situations when the physics problem is
            demanding not necessarily in terms of space and time, but in terms
            of the physics model description. Some classes of these types of
            problems are radiation transport models with complex opacities, and
            combustion models with large reaction networks. In the latter case,
            one does not expect gains from HAMR approach inside the reaction
            zone (cf.\ \cref{sec:cellular}), but this changes in the downstream
            region where multiple reaction products are advected. In that region
            the source term is essentially inactive, but the hydrodynamic
            evolution is potentially several times more expensive to compute
            because of the increased number of fluxes.  Therefore one expects
            that these classes of problems may be better suited for the
            HAMR-type approaches.

        \subsection{Aspects of implementation}
        \label{sec:implementation}

            The solver adaptive approach operates on outputs provided by the
            solver. From the implementation point of view, this requires
            communicating information between the solver and the interpolation
            operator. For example, the source term and equation of state
            interpolation both operate directly on the cell average values, and
            therefore share common MR data structures and interpolation
            routines. On the other hand, in the case of the directionally split
            hydrodynamic solver, different MR data structures and interpolation
            routines are needed, especially when the values being operated on
            are associated with cell interfaces (requiring pointwise
            interpolation). In general other types of solvers (such as
            particle-based) may require dedicated implementations of those MR
            components.

            Provided that the MR components are readily available, the main
            implementation challenge is to appropriately interface them with the
            existing solvers. This is potentially quite challenging because some
            solvers do not operate only on local data, but require additional
            information from the neighborhood of the current point. One example
            are reconstruction-evolution schemes which may use long data
            stencils. In such cases, more substantial reduction of the
            computational of the solver can be obtained by integrating the MR
            components at a lower level of the solver. This means that one may
            be required to modify the internal structures, such as individual
            loops, of those solver modules.

            In our experience with the PPM implementation available in the
            \FLASH\ code, particular components of the algorithm had to be
            converted from vector-like to scalar-like computations. Generally,
            the latter strategy is less efficient on modern computer
            architectures, but as we demonstrated the replacement of direct
            calculations with interpolation still proves beneficial in terms of
            computational cost. Alternative approaches may rely on repackaging
            the input data such that solver kernel operates on a subset of data.
            We found this approach to produce a computationally inefficient code
            due to a substantial amount of data copy operations. The version of
            the PPM solver accompanying this paper relies on the former
            fine-grained approach which restricts calculations at the loop
            level.
\section{Summary}
\label{sec:summary}
    Harten originally developed his seminal solver adaptive multiresolution
    approach for hydrodynamics problems on uniform meshes. The novelty of the
    proposed HAMR method is the extension of Harten's scheme to MR-driven
    AMR discretizations and multi-physics applications. The new scheme addresses
    the deficiency of block-structured AMR that occurs when only a fraction of
    mesh cells in an AMR block is used to resolve the structure that triggered
    mesh refinement. In those situations the remaining mesh cells inside the
    block describe the smooth part of the flow, which makes them amenable to the
    solver adaptive approach.

    We evaluated the performance of the HAMR-based code using several test
    problems. Our findings indicate that the performance gains offered by the
    new approach mainly depend on the AMR block mesh filling factor, which
    expresses the ratio of the number of cells marked for refinement to the
    overall number of cells in the block.  This factor depends on the
    application type, and for a specific application type varies within the
    computational domain. The computational gains are the highest for the cases
    in which the solution exhibits highly localized features; smaller gains are
    expected in the case of spatially extended solution structures. This
    characteristic is not surprising and is consistent with the performance
    characteristic of AMR schemes.

    The results of the numerical experiments performed indicate that the error
    due to the solver adaptive component can be controlled adequately by using a
    separate solver adaptive tolerance. Thus, the computational gains offered by
    the solver adaptive approach are shown to be achieved while solution
    accuracy, as prespecified by the user, remains under control.

    The current work opens a number of avenues for future research. For example,
    we expect HAMR to offer increased gains with the increased problem
    complexity, and therefore it would be of interest to further study the
    dependence of the efficiency on the application type.  Also, as most high
    fidelity simulations are performed on large, parallel computer systems, the
    load balancing characteristics of AMR solvers equipped with the solver
    adaptive component are expected to change and may require optimization. The
    public version of our HAMR implementation can be used as a starting point
    for these kinds of studies.
\section*{Acknowledgments}
B.G. is grateful to the SMART Scholarship-for-Service Program for their support.
Additionally, both authors acknowledge the National Energy Research Scientific
Computing Center (NERSC) for granting them with high-performance computing
resources.
\bibliography{references}

\end{document}